\documentclass[lettersize,journal]{IEEEtran}
\usepackage{amsmath,amssymb,amsfonts}
\usepackage{algorithmic}
\usepackage{graphicx}
\usepackage{xcolor}
\usepackage{amsmath,lipsum}
\usepackage[ruled]{algorithm2e}
\usepackage{float}
\usepackage{subfigure}
\usepackage{multirow}
\usepackage{multicol}
\usepackage{cite}
\usepackage{stfloats}
\usepackage{xcolor}
\usepackage{booktabs}
\usepackage{caption}
\usepackage{textcomp}
\usepackage{arydshln}
\usepackage{makecell}
\usepackage{bm}
\usepackage[colorlinks=true, linkcolor=black, citecolor=black, urlcolor=black, pdfborder={0 0 0}]{hyperref}
\usepackage[explicit]{titlesec}

\begin{document}

\title{Goal-Oriented Semantic Communication for Wireless Video Transmission via Generative AI}

\author{Nan Li,~\IEEEmembership{Member,~IEEE,} Yansha Deng,~\IEEEmembership{Senior Member,~IEEE,} Dusit Niyato,~\IEEEmembership{Fellow,~IEEE}
        % <-this % stops a space
\thanks{N. Li, Y. Deng are with the Department
of Engineering, King's College London, strand, London
WC2R 2LS, U.K. 
(email: nan.3.li@kcl.ac.uk; yansha.deng@kcl.ac.uk).}
\thanks{D. Niyato is with the School of Computer Science and Engineering, Nanyang Technological University, Singapore (email: dniyato@ntu.edu.sg).}
}

\maketitle

\begin{abstract}
Efficient video transmission is essential for seamless communication and collaboration within the visually-driven digital landscape. To achieve low latency and high-quality video transmission over a bandwidth-constrained noisy wireless channel, we propose a stable diffusion (SD)-based goal-oriented semantic communication (GSC) framework. In this framework, we first design a semantic encoder that effectively identify the keyframes from video and extract the relevant semantic information (SI) to reduce the transmission data size. We then develop a semantic decoder to reconstruct the keyframes from the received SI and further generate the full video from the reconstructed keyframes using frame interpolation to ensure high-quality reconstruction. Recognizing the impact of wireless channel noise on SI transmission, we also propose an SD-based denoiser for GSC (SD-GSC) condition on an instantaneous channel gain to remove the channel noise from the received noisy SI under a known channel. For scenarios with an unknown channel, we further propose a parallel SD denoiser for GSC (PSD-GSC) to jointly learn the distribution of channel gains and denoise the received SI. \textcolor{black}{It is shown that, with the known channel, our proposed SD-GSC outperforms state-of-the-art ADJSCC, Latent-Diff DNSC, DeepWiVe and DVST, improving Peak Signal-to-Noise Ratio (PSNR) by 69\%, 58\%, 33\% and 38\%, reducing mean squared error (MSE) by 52\%, 50\%, 41\% and 45\%, and reducing Fréchet Video Distance (FVD) by 38\%, 32\%, 22\% and 24\%, respectively. With the unknown channel, our PSD-GSC achieves a 17\% improvement in PSNR, a 29\% reduction in MSE, and a 19\% reduction in FVD compared to MMSE equalizer-enhanced SD-GSC. These significant performance improvements demonstrate the robustness and superiority of our proposed methods in enhancing video transmission quality and efficiency under various channel conditions.}
\end{abstract}


\section{Introduction}
\IEEEPARstart{T}{he} surging popularity of smart devices, coupled with the growing reliance on smart visual applications, has dramatically driven the demand for seamless, high-quality video transmission \cite{Wang_2023_IoT}. However, this growing demand of efficient and reliable large-scale video transmission capabilities poses significant challenges to existing communication bandwidth resource \cite{Wang2024TCOM}. Traditional communication systems that transmit full bit streams based on the Shannon's  technical framework, are struggling to meet these demands within bandwidth constraints \cite{Deng2023IoT}. This necessitates novel approaches for efficient video transmission that go beyond traditional methods.

\textcolor{black}{To achieve efficient video transmission, the primary challenge lies in substantially reducing transmitted data volume while maintaining high-quality video reconstruction, particularly over bandwidth-limited wireless networks \cite{Xu_2023_Globecom}. \textit{Semantic communication} (SC) has emerged as a promising approach that is expected to revolutionize the design and development of communication systems \cite{Luo_2022_WC, Kountouris_2021_MCOM}. Unlike traditional communication systems, the core principle of SC is to understand and convey the underlying meaning or intent of the message, rather than transmitting all the bits \cite{Dai_2023_WC}. This paradigm shift enables communication systems to reduce redundancy and irrelevant information, leading to more efficient data transmission. To better capture semantic information (SI) relevant to specific tasks, \textit{goal-oriented semantic communication} (GSC) framework was introduced to incorporate both semantic level and task-specific effectiveness level \cite{strinati2024goal,Strinati2021_6G}. The GSC focused on the semantic content of the message as well as its relevance and effectiveness in achieving a specific goal by dynamically prioritizing the SI transmission based on their importance to the task at hand \cite{zhou2022task, wu2023task}. A deep neural network (DNN)-based joint source-channel coding (JSCC) scheme was first developed in \cite{Bourtsoulatze_2019_TCCN} for high-resolution image transmission over additive white Gaussian noise (AWGN) and Rayleigh fading channels. Building on this foundation, advanced JSCC schemes were proposed to optimize bandwidth-agile visual data transmission \cite{Kurka2021TWC, Jiang2023JSAC}. To evaluate semantic importance, \cite{Nan2023ICC} and \cite{Huang_2023_JSAC} introduced attention mechanism-based methods to allocate varying attention weights to features, enabling the extraction of task-relevant SI for achieving the specific goals.}

\textcolor{black}{Extending these advancements to video transmission, recent works have explored strategies to efficiently extract SI while addressing the unique challenges of wireless video transmission. One such approach is shot boundary detection \cite{Zhu2023AutoShot}, which has been widely applied to tasks such as video summarization and scene detection. However, while effective for capturing broader temporal context, the SI is not strongly relevant to the goal of efficient video transmission. Following \cite{Nan2023ICC} and \cite{Huang_2023_JSAC}, \cite{Nan2023GC} extended the attention mechanism to a video recognition task by designing a 3D attention map to evaluate the importance of different pixels to achieve high inference accuracy. However, this scheme ignored the inter-frame temporal correlations that are crucial for effective video reconstruction, potentially causing redundant information transmission among similar frames. In \cite{Deepwive2022jsac}, the first end-to-end JSCC-based video transmission scheme, DeepWiVe, divides video frames into fixed Groups of Pictures (GOPs) with four frames each. The first frame in each GOP is transmitted as a keyframe using JSCC, while the remaining frames transmit motion-compensated differences relative to adjacent keyframes, also using JSCC with distinct parameters. Similarly, \cite{Wang2023JSAC} proposed a deep video semantic transmission (DVST) scheme that employs a comparable GOP structure, integrating a temporal adaptive entropy model with an Artificial Neural Network (ANN)-based nonlinear transform and conditional coding architecture to extract SI from video frames. Although these approaches improve bandwidth efficiency, their static GOP structure lacks flexibility, as the predefined keyframes may not adapt to variations in video content or bandwidth conditions. For example, the predefined keyframes within GOPs may exhibit only minor differences, resulting in redundant transmissions and suboptimal efficiency. Additionally, the end-to-end design of source coding and channel coding in these JSCC-based schemes have limitation in its adaptability, due to the fact that the end-to-end training needs to be performed for each task/goal with the lack of flexibility in the plug-and-play functionality of each module.}

It is also important to note that, the aforementioned works shared a common assumption of channel consistency between training and inference phases, breaking this assumption may lead to significant performance degradation under dynamic real-world channel conditions. The impact of channel noise is particularly critical in video transmission, where even minor distortions can significantly affect the visual quality and semantic interpretation of the content \cite{Wang2023JSAC}.
To tackle this issue, \cite{Xu_2022_TCSVT} proposed an attention mechanism-based deep JSCC (ADJSCC) scheme to dynamically adjust the signal-to-noise ratio (SNR) during training to adapt to fluctuated channel conditions and mitigate channel noise. However, this scheme remains to be end-to-end design of source coding and channel coding, limiting its flexibility to independently optimize and replace components for better adaptation to dynamic wireless channels, and goals/ tasks. Therefore, more flexible and robust modular architectures are needed to allow independent optimization and replacement of components to adapt to dynamic wireless channel conditions.

\textcolor{black}{Recent advancements of generative models, such as Generative adversarial network (GAN) \cite{NIPS2014_5ca3e9b1} and Denoising Diffusion Probabilistic Models (DDPM) \cite{chung2023diffusion}, offer new possibilities for addressing adaptive denoising challenge. Compared to GAN, DDPM has demonstrated its remarkable capability in denoising imperfect input data across various generation tasks, such as image generation \cite{Rombach_2022_CVPR} and video generation \cite{Jiang_2023_ICCV}, consistently generating high-quality images with complex textures, fine details, and sharp edges without the risk of mode collapse \cite{chung2023diffusion}. Inspired by DDPMs, two novel plug-and-play generative AI modules, channel denoising diffusion model (CDDM) \cite{Wu2023GC} and Latent diffusion denoising SC (Latent-Diff DNSC) \cite{Xu_2023_Globecom}, were proposed to characterize noisy channels for efficient image transmission. These methods highlight the effectiveness of DDPM in removing noise from received noisy SI. However, the uncontrollable generation process of DDPM makes these schemes challenging to perfectly reconstruct the details of the original image, such as appearance, style, and color \cite{Rombach_2022_CVPR}. Moreover, these schemes relies on predefined channel models (e.g., AWGN and Rayleigh fading) and assumes perfect channel state information (CSI), which limits its robustness in dynamic or imperfect channel environments.}

To fill the gap, we propose a stable diffusion (SD) model-based GSC framework for efficient video transmission over wireless fading channels, with the goal of achieving high-quality video reconstruction under latency constraints. Our main contributions are summarized as follows:
\begin{itemize}
    \item \textcolor{black}{We propose a modular SD-based GSC framework with a semantic encoder at the transmitter for extracting SI relevant to high-quality video reconstruction, and a semantic denoiser and a semantic decoder at the receiver for noise mitigation and video reconstruction, respectively.}
    \item \textcolor{black}{We design a DNN-based feature extraction module and a latency-aware keyframe selection module, to dynamically identifies keyframes with significant motion changes and extracts semantic differences between consecutive keyframes, ensuring efficient transmission of the most relevant information while reducing redundancy.}
    \item \textcolor{black}{For a known channel, we introduce an SD-based semantic denoiser for GSC (SD-GSC) that treats instantaneous channel gain as an input condition of SD to identify and remove noise. For an unknown channel, we propose a parallel SD-based semantic denoiser for GSC (PSD-GSC) that jointly estimates the instantaneous channel gain and denoises the received noisy SI.}
    
    \item \textcolor{black}{We design a DNN-based semantic reconstruction module to accurately reconstruct keyframes from the denoised SI, and a motion-appearance interpolation module to generate non-keyframes by seamlessly interpolating between reconstructed keyframes, enabling efficient and high-quality video reconstruction while preserving temporal consistency.}
\item \textcolor{black}{We conduct extensive experiments to compare our proposed methods with state-of-the-art schemes, including ADJSCC \cite{Xu_2022_TCSVT}, Latent-Diff DNSC \cite{Xu_2023_Globecom}, DeepWiVe \cite{Deepwive2022jsac}, and DVST \cite{Wang2023JSAC}, in terms of mean squared error (MSE), peak signal-to-noise ratio (PSNR), and Fréchet video distance (FVD). Under known channel conditions, SD-GSC demonstrates substantial improvements over ADJSCC, Latent-Diff DNSC, DeepWiVe, and DVST, achieving MSE reductions of approximately 52\%, 50\%, 41\%, and 45\%, respectively; PSNR improvements of 69\%, 58\%, 33\%, and 38\%; and FVD reductions of 38\%, 32\%, 22\%, and 24\%. Under unknown channel conditions, PSD-GSC further reduces MSE by approximately 29\%, improves PSNR by 17\%, and lowers FVD by 19\% compared to MMSE equalizer-enhanced SD-GSC.}
\end{itemize}

\textcolor{black}{The remainder of this article is organized as follows. The
system model and problem formulation are presented in Section \ref{section:system_model}. Section \ref{sementic_encoder} describes the designed semantic encoder for semantic information extraction. Section \ref{sementic_denoiser} details the proposed semantic denoiser for known and unknown channel conditions. Section \ref{semantic_decoder} discusses the developed semantic decoder for video construction. Section \ref{training} introduces the training pipline of the proposed framework.
The simulation results are presented and discussed in Section \ref{Performance}, and the conclusions are drawn in Section \ref{Conclusion}. The notations used in the paper are list in Table \ref{tab:notations}. }

\renewcommand\arraystretch{1.2}
\begin{table}[t]
    \centering
    \caption{\textcolor{black}{Summary of Notations}}
	\label{notations}
    \begin{tabular}{cl}
    \toprule
    \textbf{\textcolor{black}{Notation}} & \textbf{\textcolor{black}{Description}} \\
    \toprule
    \textcolor{black}{$\bm{x}$}    & \textcolor{black}{Input video with dimension $\mathbb{R}^{\mathrm{F} \times \mathrm{H} \times \mathrm{W} \times \mathrm{C}}$} \\ 
    \textcolor{black}{$\bm{x}'$}    & \textcolor{black}{Feature of Video frames with dimension $\mathbb{R}^{\mathrm{F} \times \mathrm{d}}$} \\ 
    \textcolor{black}{$\bm{z}$}    & \textcolor{black}{Semantic information of Keyframes} \\ 
    \textcolor{black}{$\bm{z}'$}    & \textcolor{black}{Noisy semantic information at the receiver} \\ 
    \textcolor{black}{$\tilde{\bm{z}}$}    & \textcolor{black}{Denoised semantic information} \\ 
    \textcolor{black}{$\hat{\bm{x}}$}    & \textcolor{black}{Reconstructed keyframes} \\ 
    \textcolor{black}{$\tilde{\bm{x}}$}    & \textcolor{black}{Generated video data} \\
    \textcolor{black}{$I$} & \textcolor{black}{Set of keyframes with $d_i$ for the $i$-th keyframe} \\

    \specialrule{0em}{2pt}{2pt}

    \textcolor{black}{$\mathcal{E}_{\texttt{fe}}$}    & \textcolor{black}{Feature extraction operation}   \\
    \textcolor{black}{$\mathcal{E}_{\texttt{ks}}$}    & \textcolor{black}{Keyframe selection operation}  \\
    \textcolor{black}{$\mathcal{F}_{\texttt{sd}}$}    & \textcolor{black}{Semantic denoising operation}  \\
    \textcolor{black}{$\mathcal{D}_{\texttt{sr}}$}    & \textcolor{black}{Semantic reconstruction operation}  \\
    \textcolor{black}{$\mathcal{D}_{\texttt{sr}}$}    & \textcolor{black}{Frame interpolation operation}  \\

    \specialrule{0em}{2pt}{2pt}

    \textcolor{black}{$\bm{h}$}    & \textcolor{black}{Channel gain}  \\
    \textcolor{black}{$\bm{n}$}    & \textcolor{black}{Noise}  \\
    \textcolor{black}{$B$}    & \textcolor{black}{Transmission bandwidth}  \\
    \textcolor{black}{$p$}    & \textcolor{black}{Transmission power}  \\
    \textcolor{black}{$t_\texttt{com}$}    & \textcolor{black}{Communication time}  \\
    \textcolor{black}{$T_\texttt{exe}$}    & \textcolor{black}{Task execution time}   \\

    \textcolor{black}{$T_\texttt{max}$}    & \textcolor{black}{Latency requirement}   \\

    \specialrule{0em}{2pt}{2pt}

    \textcolor{black}{$\bm{z}^t_1$}    & \textcolor{black}{Intermediate latent variable at diffusion step $t$} \\ 
    \textcolor{black}{$\beta_t$} & \textcolor{black}{Noise scheduling coefficient in diffusion model} \\
    \textcolor{black}{$\bm{\hat{h}}$} & \textcolor{black}{Estimated channel gain} \\  
    \textcolor{black}{$\epsilon_\theta(\bm{z}^t_1, t)$} & \textcolor{black}{Learned noise estimator of SI $\bm{z}_1$ at step $t$} \\ 
    \textcolor{black}{$\epsilon_\vartheta(\bm{h}_t, t)$} & \textcolor{black}{Learned noise estimator of channel gain $\bm{h}$ at step $t$} \\ 
    \textcolor{black}{$\bm{s}_\theta(\bm{z}^t_1, t)$} & \textcolor{black}{Score network of SI $\bm{z}_1$ at step $t$} \\ 
    \textcolor{black}{$\bm{s}_{\vartheta} \left(\boldsymbol{\bm{h}}_t, t\right)$} & \textcolor{black}{Score network of channel gain $\bm{h}$ at step $t$} \\

    \bottomrule
\end{tabular}
\label{tab:notations}
\end{table}

 \begin{figure*}
    \centering
    \includegraphics[width = \textwidth]{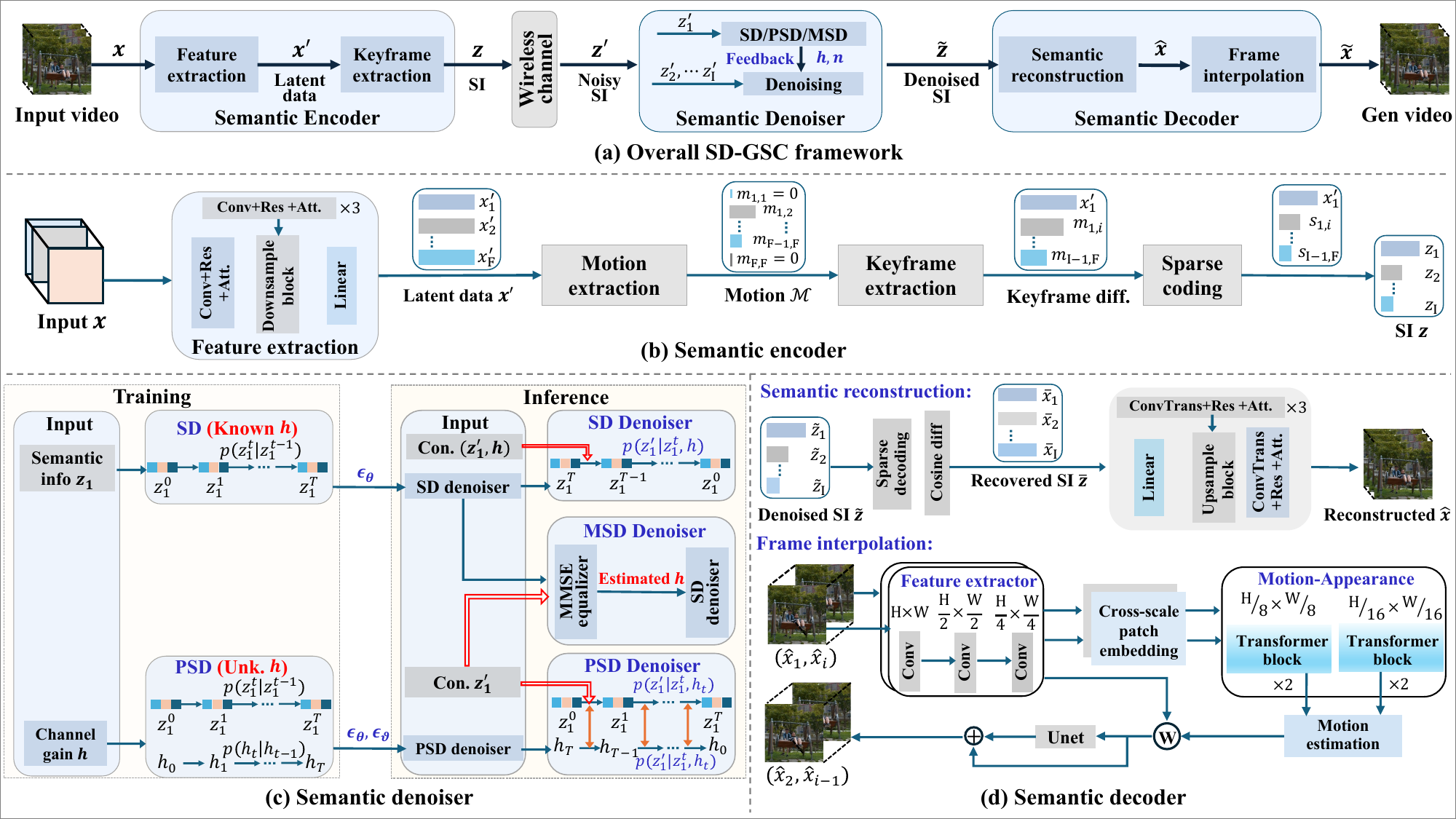}
    \caption{\textcolor{black}{Stable diffusion-based GSC framework for wireless video transmission. (a) depicts the overall framework of SD-GSC; (b) shows the detailed design of semantic encoder including the feature extraction and keyframe extraction modules; (c) displays the semantic denoiser design with SD denoiser and PSD/MSD denoisers for known and unknown channel conditions, respectively; and (d) illustrates the semantic decoder design with semantic reconstruction and frame interpolation modules. }}
    \label{fig:fig1}
\end{figure*}

\section{System Model and Problem Formulation}\label{section:system_model}
In this section, we first present our stable diffusion (SD)-based goal-oriented semantic communication (GSC) framework for wireless video transmission. Then, we formulate the optimization problem to achieve the goal of efficient video transmission.
\subsection{SD-based GSC Framework}
We consider a video transmission task between a transmitter and a receiver under Rayleigh fading channels. The communication goal is to achieve high-quality video
transmission under latency constraints. As such, we introduce the SD-based GSC framework for efficient wireless video transmission, as illustrated in Fig. \ref{fig:fig1}. The proposed framework comprises a semantic encoder at the transmitter, coupled with a semantic denoiser and semantic decoder at the receiver. The semantic encoder extracts semantic information (SI) by identifying keyframes with significant motion changes and capturing their relevant semantic representations from the video; the semantic decoder reconstructs the keyframes from the received SI accordingly, and then generates the complete video by interpolating between the reconstructed keyframes; the semantic denoiser removes wireless channel noise from the received SI to enhance the reconstruction capacity of the semantic decoder. \textcolor{black}{Specifically, for known channel conditions where the channel gain is available, we propose an \textit{SD denoiser} that effectively removes noise from the received noisy SI by conditioning on the channel gain. For scenarios with unknown channel conditions, we introduce two following denoisers:
\begin{itemize}
    \item MMSE equalizer-enhanced SD (MSD) denoiser: An MMSE equalizer estimates the instantaneous channel gain, which is then fed into the SD denoiser.
    \item Parallel SD (PSD) denoiser: Two parallel SD modules jointly learn the instantaneous channel gain and mitigate channel noise through a parallel diffusion process that simultaneously estimates channel characteristics and performs denoising.
\end{itemize}}
The details of semantic encoder, semantic denoiser, and semantic decoder are presented one by one in the following.

\subsubsection{Transmitter}
The semantic encoder at the transmitter includes the feature extraction and keyframe selection modules. 
First, the input video is represented as a 4D tensor $\bm{x} \in \mathbb{R}^{\mathrm{F} \times \mathrm{H} \times \mathrm{W} \times \mathrm{C}}$, where $\text{F}$ denotes the number of frames, $\text{C}$ represents the number of color channels, and $\text{H}$ and $\text{W}$ represent the height and width of each frame, respectively. Initially, the high-dimensional input video $\bm{x}$ is fed into the feature extraction module to extract the most critical and semantically relevant information related to the communication goal from each frame and outputs a lower-dimensional latent tensor $\bm{x}' \in \mathbb{R}^{\mathrm{F} \times \mathrm{d}}$, with $\mathrm{d} \ll \mathrm{H} \times \mathrm{W} \times \mathrm{C}$. Mathematically, we express this video feature extraction process as 
\begin{equation}
    \bm{x}' = \mathcal{E}_{\texttt{fe}} (\bm{x}),
\end{equation}
where $\mathcal{E}_{\texttt{fe}}(\cdot)$ denotes the operation of feature extraction.

The latent tensor $\bm{x}'$ is then processed by the keyframe selection module to identify the semantically significant frames, known as keyframes. The resulting SI of these selected keyframes is represented as a tensor \(\bm{z} = \{ z_i \mid z_i \in \mathbb{R}^{1 \times d_i}, \forall i \in \mathcal{I} \}\), where $\mathcal{I}$ denotes the set of selected keyframes. The SI $\bm{z}$ is mathematically defined as
\begin{equation}
\bm{z} = \mathcal{E}_{\texttt{ks}}(\bm{x}'),
\end{equation}
where $\mathcal{E}_{\texttt{ks}}(\cdot)$ represents the keyframe selection operation.
\subsubsection{Wireless Channel}
The SI \(\bm{z}\) is then transmitted over a wireless channel under Rayleigh fading, which introduces impairments such as signal attenuation, multipath fading, and additive noise. The received noisy SI of keyframes, \(\bm{z}'\), can be represented as
\begin{equation}
\bm{z}' = \bm{h} \cdot \bm{z} + \bm{n},
\end{equation}
where \(\bm{h}\) represents the channel gain between the transmitter and receiver, and \(\bm{n} \sim \mathcal{CN}(0, \sigma^2)\) is the additive white Gaussian noise (AWGN) with noise power \(\sigma^2\).

Let us denote the transmission bandwidth as \(B\) and the transmission power as \(p\), the transmission data rate can be calculated as 
\begin{equation}
R = B \log_2 \left(1 + \frac{p\bm{h}}{\bm{n} }\right).
\end{equation}

Given that the semantic tensor \(\bm{z}\) is typically in float32 data type in machine learning frameworks such as PyTorch and TensorFlow \cite{NEURIPS2019_bdbca288}, the communication time can be calculated as
\begin{equation}
t_\texttt{com} = \frac{\sum_{i=1}^{\mathrm{I}} 32 d_i}{R},
\end{equation}
where $\mathrm{I}$ is the total number of selected keyframes, and $d_i$ is the dimension of the SI $z_i$ for the $i$th keyframe.

\subsubsection{Receiver}
At the receiver, the primary objective is to reconstruct the original video from the received SI while mitigating the impact of channel-induced noise and distortions. The semantic decoder at the receiver includes the semantic denoiser, semantic reconstruction, and frame interpolation modules.

The received noisy SI ${\bm{z}'}$ is first processed by the semantic denoiser module. The semantic denoiser progressively denoises the noisy SI ${\bm{z}'}$ to obtain a denoised semantic vector $\tilde{\bm{z}}$ as
\begin{equation}
\tilde{\bm{z}} = \mathcal{F}_{\texttt{sd}}({\bm{z}'}),
\end{equation}
where $\mathcal{F}_{\texttt{sd}}(\cdot)$ is the semantic denosing operation. 

The denoised semantic representation  ${\tilde{\bm{z}}}$ is then passed to the semantic reconstruction module to reconstruct the keyframes ${\hat{\bm{x}}}$ by learning the inverse mapping from the latent space to the high-dimensional pixel space. The reconstructed keyframes ${\hat{\bm{x}}}$ can be denoted as 
\begin{equation}
   \hat{ \bm{x}} =  \mathcal{D}_{\texttt{sr}}(\tilde{\bm{z}}),
\end{equation}
where $ \mathcal{D}_{\texttt{sr}}(\cdot)$ denotes the semantic reconstruction operation.

Finally, the reconstructed keyframes $\hat { \bm{x}} $ are fed into the frame interpolation module to generate the complete video $\tilde{\bm{x}}$ by interpolating between the reconstructed keyframes ${\hat{\bm{x}}}$. The generated video data $\tilde{\bm{x}}$ can be expressed as
\begin{equation}
   \tilde{ \bm{x}} =  \mathcal{D}_\texttt{fi} (\hat{\bm{x}}),
\end{equation}
where $\mathcal{D}_\texttt{fi}  (\cdot)$ represents the frame interpolation operation.

\subsection{Problem Formulation}
To perform video transmission via a wireless channel, the execution time includes the computation time for semantic encoding at the transmitter, the communication time for transmitting SI, and the computation time for semantic denoising and semantic decoding at the receiver. Therefore, the execution time can be expressed as
\begin{equation}
    T_{\texttt{exe}} =   \underbrace{t_\texttt{fe} +t_\texttt{ks} }_{\text{Encoding time}} + t_\text{com}+ \underbrace{t_\texttt{sd} +t_\texttt{sr} +t_\texttt{fi} }_{\text{Decoding time}},
\end{equation}
where $t_\texttt{fe}$ and $t_\texttt{ks}$ represent the measured computation time of feature extraction and keyframe selection modules at the transmitter, respectively; $t_\texttt{sd}$, $t_\texttt{sr}$, and $t_\texttt{fi}$ denote the measured computation time of semantic denoising, semantic reconstruction, and frame interpolation at the receiver, respectively.

The primary objective of the wireless video transmission scheme is to accurately reconstruct the original video at the receiver while minimizing the distortion introduced by the wireless channel. To quantify the reconstruction quality, we introduce the mean squared error (MSE) as the distortion metric, which measures the average squared difference between the transmitted video frames at the transmitter and the reconstructed video frames at the receiver. The objective is to minimize the average MSE between the original video frames and the reconstructed video frames within the time constraints as 
\begin{align}
    & \min_{\mathcal{P}} \frac{1}{N} \sum_{\bm{x} \in \bf{X}} \big|\big|\bm{x} - \tilde{\bm{x}}\big|\big|^2 \nonumber \\
    &  \;\; \;\textrm{s.t.}  \;\; T_{\texttt{exe}} \leq T_\texttt{max}, 
    \label{eq:obj}
\end{align} 
where $N$ is the size of dataset $\bf{X}$; $\mathcal{P}$ represents the set of learnable parameters in the various components of the proposed system, including the feature extraction, keyframe selection, semantic denoising, semantic reconstruction, and frame interpolation; $T_\texttt{max}$ is the latency requirement of the video transmission task.

\begin{figure*}[t]
    \centering
    \includegraphics[width = \textwidth]{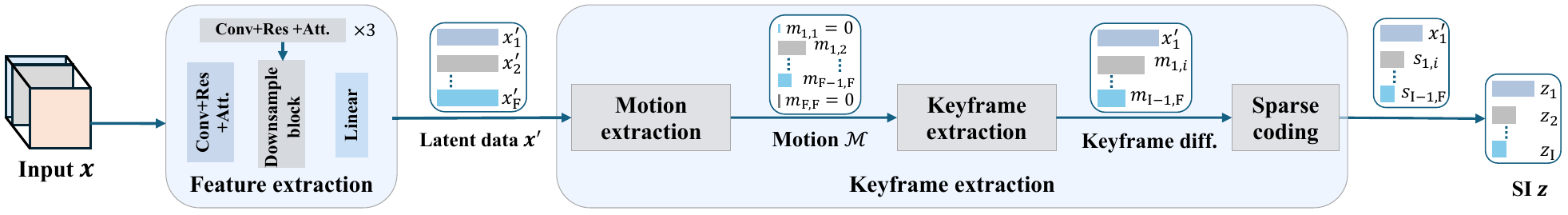}
    \caption{\textcolor{black}{Semantic encoder with feature extraction and keyframe extraction. }}
\label{Fig:semantic_encoder}
\end{figure*}
\section{Semantic Encoder}\label{sementic_encoder}
In this section, we present the details of the semantic encoder at the transmitter, starting with the design of the feature extraction module and then introducing the keyframe selection module.

\subsection{Feature Extraction}
To extract deep feature information for identifying keyframes while minimizing data transmission requirements and optimizing video reconstruction quality at the receiver, we propose a deep neural network-based feature extraction module, as shown in Fig. \ref{Fig:semantic_encoder}. This module encodes input video into a low-dimensional latent space to efficiently preserve the most critical feature information and discard redundant information for keyframe extraction and high-quality reconstruction. 

The network architecture is designed using a similar downsample framework as U-Net \cite{Unet}, and the details are presented as follows. The input video \(\bm{x} \) is fed into an initial convolutional layer \texttt{\( (3 \times 3 \,\, \text{kernel}, \text{stride} \,\, 1, \text{padding} \,\, 1) \)}, followed by a residual block and an attention block, both operating on the same channel dimension \(\mathrm{C}=3\) (i.e., the RGB channels), to capture and preserve essential features. The output tensor is then processed by three successive downsampling stages, each consisting of a convolutional layer  \texttt{\( (3 \times 3 \, \,\text{kernel}, \text{stride} \,\, 2, \text{padding} \,\, 1) \)} to halve the spatial dimensions, a residual block for feature refinement, and an attention block to focus on the most relevant spatial information. These stages operate at progressively increasing channel dimensions \(c \in \{64, 128, 256, 512\}\), allowing the network to capture more abstract and complex semantic attributes while reducing the spatial dimensions. After that, batch normalization is applied to stabilize the learning process, followed by a ReLU activation function to introduce non-linearity. Finally, the resulting tensor are embedded into a latent representation \(\bm{x}' \in \mathbb{R}^{\mathrm{F}\times \mathrm{d}}, \mathrm{d} = 4096\) using a linear layer.

\begin{algorithm}[t]
\renewcommand{\algorithmicrequire}{\textbf{Input:}}
\renewcommand{\algorithmicensure}{\textbf{Output:}}
    \caption{Keyframe Selection} 
    \label{alg1}
    \begin{algorithmic}[1]
    \REQUIRE Video \(\bm{x}\), latency requirement \(T_\texttt{max}\), transmission power \(p\), bandwidth \(B\),  channel gain \(h\), noise power \(\sigma^2\)
    \ENSURE  The set of keyframes \(\mathcal{I}\)
    
    \STATE Initialize the set of keyframes \(\mathcal{I} = \varnothing\).
    \STATE Extract the SI from \(\bm{x}\) using (1).
    \STATE Calculate the cosine difference \(\bm{m}_{i, j}\) between each pair of frames using (\ref{eq:cosine}). 
    \STATE Apply sparse coding to represent \(\bm{m}_{i, j}\) as \(\bm{s}_{i, j}\).
    \STATE Add the initial frame \(1\) and the last frame \(F\) to $\mathcal{I} = \mathcal{I} \cup \{1, F\}$.
    \STATE Calculate the total completion time \(\mathcal{T}_0\) for $\mathcal{I}$.
    \STATE Initialize a max-heap \(H\) with tuples \((\sum_{j \in \mathcal{I}} \bm{s}_{i,j}, i \notin \mathcal{I})\).
    \WHILE{$\mathcal{T}_0 < T_\texttt{max}$}
        \STATE Select frame \(i^*\) with the maximum change from \(H\).
        \STATE Set a temporal set of keyframes $\mathcal{I}' = \mathcal{I} \cup \{i^*\}$.
        \STATE Calculate the total completion time \(\mathcal{T}'\) for $\mathcal{I}'$.
        \IF {$\mathcal{T}' < T_\texttt{max}$}
            \STATE $\mathcal{I} = \mathcal{I}'$, \(\mathcal{T}_0 = \mathcal{T}'\).
            \STATE Update the max-heap \(H\) involving \(i^*\).
        \ENDIF
    \ENDWHILE
    \end{algorithmic}
\end{algorithm}
\subsection{Keyframes Selection}
\textcolor{black}{Video sequences frequently encompass redundant frames characterized by minimal visual variation between consecutive images. For example, in a football match recording, frames displaying high-speed action, such as a player shooting at the goal or a goalkeeper diving for a save, show considerable visual differences compared to frames with little movement like players walking across the field or remaining stationary.} To identify the most informative and representative frames (i.e., keyframes) in the input video, we propose a keyframe selection module that analyzes temporal correlation by calculating inter-frame differences and detecting substantial visual changes, as shown in Fig. \ref{Fig:semantic_encoder}. Specifically, we quantify these differences using the cosine similarity between the feature representations \(\bm{x}'_i\) and \(\bm{x}'_j\) of frames \(\bm{x}_i\) and \(\bm{x}_j\) as 
\begin{equation}
\bm{m}_{i, j} = 1 - \frac{\bm{x}'_i \cdot \bm{x}'_j}{\|\mathbf{x}'_i\| \, \|\mathbf{x}'_j\|},
\label{eq:cosine}
\end{equation}
where a larger $\bm{m}_{i,j}$ indicates greater visual dissimilarity between the frame pair, suggesting that one of these frames could be a potential keyframe. 

During video transmission, we start by sending the complete feature representation of the first keyframe,
\(\bm{x}'_1\), to build a knowledge base for future frame recovery. For subsequent keyframes, instead of transmitting entire feature representation, we send a sparse tensor representation \(\bm{s}_{i,j}\) of the cosine differences \(\bm{m}_{i,j}\) between consecutive keyframes. This approach leverages the fact that differences between adjacent keyframes are often sparse (i.e., has many zero values), as most of the visual information remains unchanged between consecutive keyframes. By exploiting this sparsity, we can remove the redundant information thereby further reducing the bandwidth requirements. The keyframes selection is determined by meeting specified latency constraints while ensuing high quality reconstruction during transmission. The details of keyframe selection module are presented in \textbf{Algorithm 1}. Initially, since the frame interpolation module in the semantic decoder requires at least two frames, the first and last frames are selected as the keyframes. Then, additional frames that have the maximum cosine difference from the existing keyframes in \(\mathcal{I}\) are iteratively included to obtain the optimal keyframes set ${\mathcal{I}} = \left\{1, 2, \ldots, \mathrm{I} \right\}$ until the execution time of the video task exceeds the latency requirement.

\section{Semantic Denoiser}\label{sementic_denoiser}
In this section, we first propose an \textit{SD denoiser} to effectively eliminate the effect of wireless channel noise on SI transmission under known channel scenario, and then propose a \textit{PSD denoiser} to handle the unknown channel scenario.

\subsection{SD Denoiser under Known Channel}
\begin{figure}[t]
    \centering
    \includegraphics[width =  0.5\textwidth]{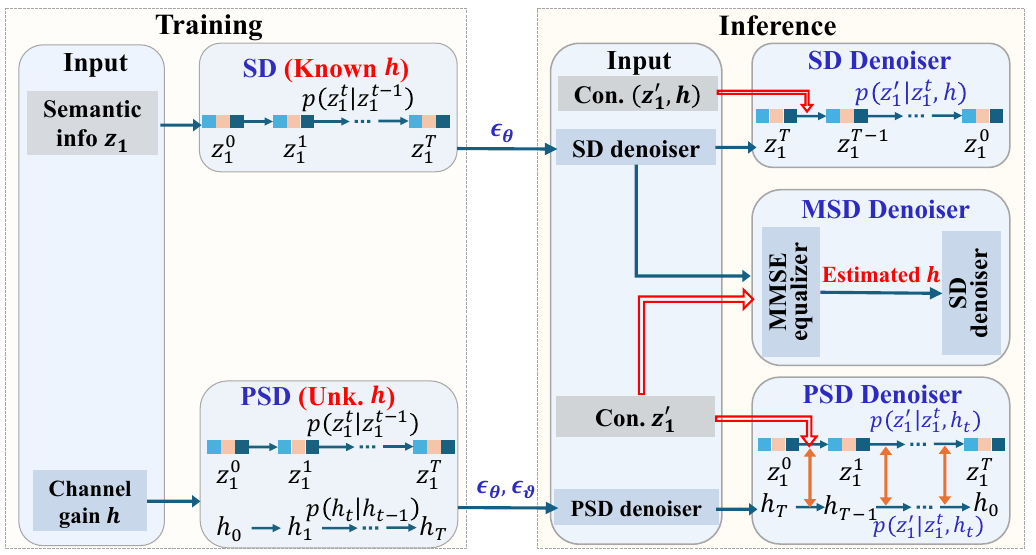}
    \caption{\textcolor{black}{Semantic denoiser under known and unkown channels. }}
\label{Fig:semantic_denoiser}
\end{figure}

To handle the wireless channel noise during SI transmission, the uncontrolled image generation process of DDPM, as applied in Latent-Diff DNSC \cite{Xu_2023_Globecom}, potentially degrades the reconstruction quality \cite{Rombach_2022_CVPR}. To overcome this challenge, we propose the \textit{SD denoiser} that uses the received noisy SI and instantaneous channel gain as control conditions to effectively remove channel noise, as shown in Fig. \ref{Fig:semantic_denoiser}. Note that, due to the computation-intensive nature of the \textit{SD denoiser}, we only apply it to denoise the received noisy SI of the first keyframe, and then use the noise information $\bm{n}$ captured by \textit{SD denoiser} to denoise the subsequent keyframes.

\subsubsection{Forward Process of DDPM}
The forward process (i.e., training process) is applied to the transmitted SI of the first keyframe, \(\bm{z}_1\), generated by the semantic encoder. This involves iteratively adding Gaussian noise to the initial distribution \(\bm{z}_1^0 \sim p(\bm{z}_1)\) over \(\mathrm{T}\) time steps, gradually approaching an isotropic Gaussian distribution \(\bm{z}_1^\mathrm{T} \sim \mathcal{N}(0, \mathbf{I})\). This process can be viewed as a discrete-time Markov chain, where the current state $\bm{z}_1^t$ is obtained by adding noise to the previous state $\bm{z}_1^{t-1}$. In DDPM \cite{NEURIPS2020_4c5bcfec}, at time step $t \in [0, \mathrm{T}] $, the forward process of $\bm{z}_1^t $ is expressed as 
\begin{equation}
\bm{z}_1^{t}=\sqrt{1-\beta_{t}} \bm{z}_1^{t-1}+\sqrt{\beta_{t}} \epsilon,
\label{eq:forward}
\end{equation}
where $\beta_{t} \in (0,1)$ is the noise scheduling function, typically modeled as a monotonically increasing linear function of $t$, and $\epsilon \sim \mathcal{N}(0, \mathbf{I})$. 

\textcolor{black}{From the score-based perspective (i.e., the gradient of the log probability density with respect to the data $\bm{z}_1^t$ at each noise scale $\beta_t$), the forward stochastic differentiable equation (SDE) can be expressed as
\begin{align}
\mathrm{d}^{(f)} \bm{z}_1^t &= - \frac{\beta_t}{2} \bm{z}_1^t \, \mathrm{d}t + \sqrt{\beta_t} \, \mathrm{d} w_t \nonumber \\ 
& = f(\bm{z}_1^t, t) \, \mathrm{d} t + g(t) \, \mathrm{d} w_t,
\label{eq:addnoise}
\end{align}
where $f(\bm{z}_t, t) = - \frac{\beta_t}{2} \bm{z}_t$ is the drift term, $g(t) = \sqrt{\beta_t}$ is the diffusion coefficient, $\mathrm{d} w_t$ is the standard Wiener process.}

\textcolor{black}{
\subsubsection{Reverse Process of DDPM}
The reverse process (i.e., inference process) of DDPM aims to recover the original SI \(\bm{z}_1\) from the noisy sample $\bm{z}_1^\mathrm{T} \sim \mathcal{N}(0, \mathbf{I})$. This process mirrors the marginal distribution of the forward process \(p(\bm{z}_1^t)\) but with the drift direction reversed. Correspondingly, the reverse-time SDE becomes 
\begin{align}
    \mathrm{d}^{(r)}\bm{z}_1^t &= \big[f(\bm{z}_1^t, t) - g(t)^2 \nabla_{\bm{z}_1^t} \log p(\bm{z}_1^t)\big] \mathrm{d}t + g(t) \, \mathrm{d}\overline{w}_t \nonumber \\
    & = \big [ - \frac{\beta_t}{2} \bm{z}_1^t - \beta_t \nabla_{\bm{z}_1^t} \log p(\bm{z}_1^t) \big ] \mathrm{d} t + \sqrt{\beta_t} \, \mathrm{d}\overline{w}_t,
    \label{eq:reverse}
\end{align}
where $s(\bm{z}_1^t, t) = \nabla_{\bm{z}_1^t} \log p(\bm{z}_1^t)$ is the score function, which is intractable and needs to be approximated using a neural network $s_\theta(\bm{z}_1^t, t)$. }

Since $\nabla_{\bm{z}_1^t} \log p (\bm{z}_1^t) = \nabla_{\bm{z}_1^t} \log p\left(\bm{z}_1^t \mid \boldsymbol{\bm{z}}_0^0\right)$, we can approximate $\nabla_{\bm{z}_1^t} \log p (\bm{z}_1^t) \simeq	 \boldsymbol{s}_\theta\left(\bm{z}_1^t, t\right)$ for the reverse process in (\ref{eq:reverse}) by solving the following minimization problem during the training in the forward process \cite{Vincent2011NC}: 
\begin{equation}
\theta^* = \underset{\theta}{\operatorname{argmin}} \mathbb{E}_{\bm{z}_1^t, \bm{z}_1^0}\left[\left\|\boldsymbol{s}_\theta\left(\boldsymbol{\bm{z}}_1^t, t\right)-\nabla_{\boldsymbol{\bm{z}}_1^t} \log p\left(\boldsymbol{\bm{z}}_1^t | \boldsymbol{\bm{z}}_0^0\right)\right\|_2^2\right],
\label{eq:trainprocess}
\end{equation}
where the trained score network $\bm{s}_\theta\left(\bm{z}_1^t, t\right) $ can be denoted by using Tweedie’s identity \cite{NEURIPS2021_077b83af} as
\begin{equation}
    \bm{s}_\theta\left(\boldsymbol{\bm{z}}_1^t, t\right) = \nabla_{\bm{z}_1^t} \log p(\bm{z}_1^t)  = - \frac{1}{\sqrt{1-\bar{\alpha}_{t}}}\epsilon_\theta\left(\boldsymbol{\bm{z}}_1^t, t\right),
    \label{eq:score}
\end{equation}
where $\alpha_t = 1-\beta_t$ and $\bar{\alpha}_{t}=\prod_{i=1}^{t}\left(1-\alpha_{i}\right)$, and the parameter $\epsilon_\theta\left(\boldsymbol{\bm{z}}_1^t, t\right)$ is the learned noise estimator at time step $t$.

\subsubsection{Reverse Process of Stable Diffusion}
Evidently, the above reverse process of DDPM from a random Gaussian sample $\bm{z}_1^\mathrm{T}$ cannot ensure the quality of image reconstruction. To allow for more controllable and guided image generation, we design an \textit{SD denoiser} conditional on the received noisy SI ${\bm{z}_1'}$ and instantaneous channel gain ${\bm{h}}$. Here, the \textit{SD denoiser} follows the same forward process as DDPM \cite{Rombach_2022_CVPR} .

Leveraging the diffusion model as the prior, it is straightforward to modify (\ref{eq:reverse}) to derive the reverse process of SD from the posterior distribution
\begin{equation}
\begin{aligned}
   \mathrm{d}^{(r)} \bm{z}_1^t & = [ - \frac{\beta_{t}}{2} \bm{z}_1^t - \beta_{t} \nabla_{\bm{z}_1^t} \log p\left( \bm{z}_1^t \mid \bm{z}'_1 \right) ] \mathrm{d} t + \sqrt{\beta_{t}} \mathrm{d} \overline{w}_t \\ 
   & = [ - \frac{\beta_{t}}{2} \bm{z}_1^t - \beta_{t} (\nabla_{\bm{z}_1^t} \log p (\bm{z}_1^t)  \\
   & + \nabla_{\bm{z}_1^t} \log p(\bm{z}_1'|\bm{z}_1^t ,\bm{h}) )] \mathrm{d} t +\sqrt{\beta_{t}} \mathrm{d} \overline{w}_t ,
    \label{eq:con_reverse}
\end{aligned}
\end{equation}
where we use
\begin{align}
       \nabla_{\bm{z}_1^t} \log p\left( \bm{z}_1^t \right) &=  \nabla_{\bm{z}_1^t} \log p\left( \bm{z}_1^t \mid \bm{z}_1' \right) + \underbrace{\nabla_{\bm{z}_1^t} \log p\left( \bm{z}_1' \right)}_{0} 
        \nonumber \\
         & = \nabla_{\bm{z}_1^t} \log p\left( \bm{z}_1^t \mid \bm{z}_1' \right) ,
\end{align}
and 
\begin{align}
    \nabla_{\bm{z}_1^t} \log p\left( \bm{z}_1^t \mid \bm{z}_1' \right) = & \nabla_{\bm{z}_1^t} \log p\left( \bm{z}_1^t \right) + \nabla_{\bm{z}_1^t} \log p\left( \bm{z}_1'  \mid \bm{z}_1^t \right) \nonumber \\
    = & \nabla_{\bm{z}_1^t} \log p\left( \bm{z}_1^t \right) + \nabla_{\bm{z}_1^t} \log p\left( \bm{z}_1'  \mid \bm{z}_1^t, \bm{h} \right) \nonumber \\
    & +  \underbrace{\nabla_{\bm{z}_1^t} \log p\left( \bm{h} \right)}_{0} \nonumber \\
    = &  \nabla_{\bm{z}_1^t} \log p\left( \bm{z}_1^t \right) + \nabla_{\bm{z}_1^t} \log p\left( \bm{z}_1'  \mid \bm{z}_1^t, \bm{h} \right) .
\end{align}

By discretizing the reverse process in (\ref{eq:con_reverse}), we have
\begin{equation}
\begin{aligned}
    \bm{z}_1^{t-1} = & \frac{1}{\sqrt{\alpha_{t}} }(\bm{z}_1^{t} + \beta_t [\bm{s}_\theta\left(\boldsymbol{\bm{z}}_1^t, t\right) + \nabla_{\bm{z}_1^t} \log p(\bm{z}_1'|\bm{z}_1^t,\bm{h}) ] ) \\
    & + \sqrt{\beta_t} \mathcal{N}(0, \mathbf{I}).
\end{aligned}
\label{eq:conditionreverse}
\end{equation}

To solve the reverse process in (\ref{eq:conditionreverse}), the main challenge lies in the posterior distribution $p(\bm{z}_1'|\bm{z}_1^t,\bm{h})$. While the relationship between the received noisy SI \(\bm{z}_1'\) and the transmitted SI \(\bm{z}_1^0\) is known, the relationship between the intermediate data \(\bm{z}_1^t\) at the \(t\)th step of the forward process and \(\bm{z}_1'\) remains unknown. To tackle this issue, we express $p\left(\bm{z}_1' \mid \bm{z}_1^t\right)$ as 
\begin{equation}
    p\left(\bm{z}_1' \mid \bm{z}_1^t\right)=\int p\left(\bm{z}_1' \mid \bm{z}_1^0 \right) p\left(\bm{z}_1^0 \mid \bm{z}_1^t\right) \mathrm{d} \bm{z}_1^0,
    \label{eq:marginal}
\end{equation}
where the mean of $p(\bm{z}_1^0\mid \bm{z}_1^t)$ can be approximated by a delta function as
\begin{equation}
     p\left(\bm{z}_1^0 \mid \bm{z}_1^t\right) \simeq	 \delta_{\textbf{E}[\bm{z}_1^0 \mid \bm{z}_1^t]}(\bm{z}_1^0).
\end{equation}

To estimate the $E[\bm{z}_1^0|\bm{z}_1^t]$, we can use the well-trained noise estimator $\epsilon_{\theta}(\bm{z}_1^t, t)$ in the forward process (\ref{eq:trainprocess}) to obtain the estimation $ E[\bm{z}_1^0|\bm{z}_1^t] = \hat{\bm{z}}_1^t$ as
\begin{equation}
    \hat{\bm{z}}_1^t =\frac{1}{\sqrt{\alpha_{t}} }(\bm{z}_1^{t}- \sqrt{1-\bar{\alpha}_{t-1}}\epsilon_{\theta}(\bm{z}_1^t,t))
    \label{eq:z0}.
\end{equation}

Using (\ref{eq:z0}), the approximation $p(\bm{z}_1^0|\bm{z}_1^t)$ leads to the following formula for the gradient of the log-likelihood:
\begin{equation}
    \nabla_{\bm{z}_1^t} \log p(\bm{z}_1'|\bm{z}_1^t,\bm{h}) = - \frac{1-\bar{\alpha}_{t}}{\beta_{t}\left(1-\bar{\alpha}_{t-1}\right)} \nabla_{\bm{z}_1^t} ||\bm{z}_1'-\bm{h}  {\bm{z}_1^t}||^2.
\end{equation}

\subsection{PSD Denoiser under Unknown Channel}
\textcolor{black}{Notably, the \textit{SD denoiser} is only applicable when the instantaneous channel gain $\bm{h}$ is known, and hence cannot be directly used for the scenarios with imperfect estimation of $\bm{h}$ (e.g., SISO Rayleigh fading channels \cite{Punchihewa2011tWC}).} To solve this issue, we propose a parallel SD  (PSD) denoiser to jointly estimate the channel gain and remove the noise, as shown in Fig. \ref{Fig:semantic_denoiser}. Similar to the \textit{SD denoiser}, the \textit{PSD denoiser} is applied only to the received noisy SI of the first keyframe.
\textcolor{black}{
\subsubsection{Forward Process}
Given that $\bm{z}_1$ and $\bm{h}$ are independent, the posterior probability can be expressed as: 
\begin{equation} p(\bm{z}_1,\bm{h}|\bm{\bm{z}_1'}) \propto p(\bm{\bm{z}_1'}|\bm{z}_1,\bm{h}) p(\bm{z}_1) p(\bm{h}), 
\end{equation}
where represents the received noisy SI. Based on this, we train two separate forward processes for $\bm{z}_1$ and $\bm{h}$, respectively.}

\textcolor{black}{Similar to equations (\ref{eq:score}) and (\ref{eq:z0}), the score network of $\bm{h}$ and the estimation $E[\bm{h}_0|\bm{h}_t]= \hat{\bm{h}}_t$, can be derived using
\begin{equation}
    \bm{s}_{\vartheta} \left(\boldsymbol{\bm{h}}_t, t\right) = \nabla_{\bm{h}_t} \log p(\bm{h}_t)  = - \frac{1}{\sqrt{1-\bar{\alpha}_{t}}}\epsilon_{\vartheta}(\bm{h}_t,t),
    \label{eq:score1}
\end{equation}
and
\begin{equation}
    \hat{\bm{h}}_t =\frac{1}{\sqrt{\alpha_{t}} }(\bm{h}_{t}- \sqrt{1-\bar{\alpha}_{t-1}}\epsilon_{\vartheta}(\bm{h}_t,t))
    \label{eq:h0},
\end{equation}
where $\epsilon_\vartheta\left(\boldsymbol{\bm{h}}_t, t\right)$ represents the learned noise estimator for the channel gain at time step $t$, parameterized by $\vartheta$.}
\textcolor{black}{
\subsubsection{Reverse Process}
The reverse processes for PSD follow the framework in (\ref{eq:con_reverse}), and the two reverse processes can be expressed as
\begin{equation}
\begin{aligned}
   \mathrm{d}^{(r)} \bm{z}_1^t = & \big[ - \frac{\beta_{t}}{2} \bm{z}_1^t - \beta_{t} (\nabla_{\bm{z}_1^t} \log p (\bm{z}_1^t) \\ 
   & + \nabla_{{\bm{z}}_1^t} \log p(\bm{z}_1'|\bm{z}_1^t, \bm{h}_t) )\big] \mathrm{d} t + \sqrt{\beta_{t}} \mathrm{d} \overline{w}_t,
\end{aligned}
\end{equation}
and
\begin{equation}
\begin{aligned}
   \mathrm{d}^{(r)} \bm{h}_t = & \big[ - \frac{\beta_{t}}{2} \bm{h}_t - \beta_{t} (\nabla_{\bm{h}_t} \log p (\bm{h}_t) \\ 
   & + \nabla_{{\bm{h}}_t} \log p(\bm{z}_1'|\bm{z}_1^t, \bm{h}_t) ) \big] \mathrm{d} t + \sqrt{\beta_{t}} \mathrm{d} \overline{w}_t.
\end{aligned}
\end{equation}
Correspondingly, the above two reverse processes can be discretized in a similar way to (\ref{eq:conditionreverse}) and expressed as 
\begin{align}
    \bm{z}_1^{t-1} = & \frac{1}{\sqrt{\alpha_{t}} }\big(\bm{z}_1^{t} +  \beta_t \big[\nabla_{\bm{z}_1^t} \log p(\bm{z}_1'|\bm{z}_1^t,\bm{h}_t) +  \bm{s}_{\theta} \left(\boldsymbol{\bm{z}}_1^t, t\right) \big] \big)  
 \nonumber \\
    &  + \sqrt{\beta_t} \mathcal{N}(0, \mathbf{I}),
\end{align}
and 
\begin{align}
    \bm{h}_{t-1} =  & \frac{1}{\sqrt{\alpha_{t}} }\big(\bm{h}_{t} +  \beta_t \big[\nabla_{\bm{h}_t} \log p(\bm{z}_1'|\bm{z}_1^t,\bm{h}_t)+  \bm{s}_{\vartheta} \left(\boldsymbol{\bm{h}}_t, t\right) \big] \big) \nonumber  \\
    &  + \sqrt{\beta_t} \mathcal{N}(0, \mathbf{I}),
\end{align}
where 
\begin{equation}
    \nabla_{\bm{z}_1^t} \log p(\bm{z}_1'|\bm{z}_1^t,\bm{h}_t) = - \frac{1-\bar{\alpha}_{t}}{\beta_{t}\left(1-\bar{\alpha}_{t-1}\right)} \nabla_{\bm{z}_1^t} ||\bm{z}_1'-\bm{h}_t  {\bm{z}_1^t}||^2,
\end{equation}
and
\begin{equation}
    \nabla_{\bm{h}_t} \log p(\bm{z}_1'|\bm{z}_1^t,\bm{h}_t) = - \frac{1-\bar{\alpha}_{t}}{\beta_{t}\left(1-\bar{\alpha}_{t-1}\right)} \nabla_{\bm{h}_t} ||\bm{z}_1'-\bm{h}_t  {\bm{z}_1^t}||^2.
\end{equation}
}

\textcolor{black}{Given the sparse structure of certain wireless channels, we use $\ell_1 $ regularization to sparse the channel gain by augmenting the diffusion prior thereby 
better stabilizing the reconstruction. The estimated channel gain is then updated as
\begin{equation}
    \bm{h}_{t-1} = \bm{h}_{t-1} - \alpha (||\bm{z}_1'-\bm{h}_{t-1} * \bm{z}_1^{t-1}||_2 + \phi ||\bm{h}_{t-1}||),
\end{equation}
where $\phi$ is the regularization strength.}

\section{Semantic Decoder}\label{semantic_decoder}
This section provides the details of the semantic reconstruction module and frame interpolation module within the semantic decoder at the receiver.
\subsection{Semantic Reconstruction}
\begin{figure}[t]
    \centering
    \includegraphics[width =  0.5\textwidth]{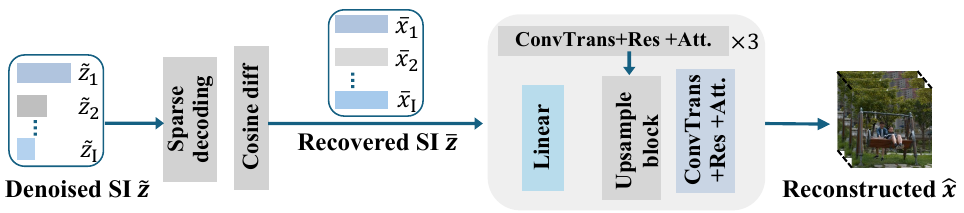}
    \caption{\textcolor{black}{Semantic reconstruction. }}
\label{Fig:semantic_reconstruction}
\end{figure}

To ensure accurate reconstruction of keyframes at the receiver, the semantic reconstruction module is designed to minimize visual artifacts and ensure that the reconstructed keyframes closely resemble the original frames in both content and quality. This is achieved by progressively reconstructing the spatial and semantic details of the keyframes from the denoised low-dimensional SI \(\tilde{\bm{z}}\), as shown in Fig. \ref{Fig:semantic_reconstruction}.

The architecture of this network follows an upsampling framework similar to that of U-Net \cite{Unet}, and is detailed as follows. It begins with sparse decoding and cosine difference to recover the latent representation of keyframes, \(\bar{\bm{z}} \in \mathbb{R}^{ \mathrm{I} \times \mathrm{d} }\), from the denoised SI \(\tilde{\bm{z}}\). Then, a linear layer is applied to transform the latent representation \(\bar{\bm{z}} \) into a higher-dimensional feature space, \( \mathbb{R}^{ \mathrm{I} \times \mathrm{h} \times \mathrm{w} \times \mathrm{c}}\). Subsequently, the upsampling unfolds across three stages, each incorporating a transposed convolutional layer \texttt{\( (3 \times 3 \,\, \text{kernel}, \text{stride} \,\, 2, \text{padding} \,\, 1) \)} to double the spatial dimensions, followed by a residual block for refining the features and an attention block to enhance the most critical spatial features, ensuring that the reconstruction captures the essential details accurately. These stages progressively reduce the tensor's channel dimensions \(c \in \{512, 256, 128, 64\}\), allowing the network to capture intricate details while expanding the spatial dimensions of the feature map. The final refinement employs a transposed convolutional layer \texttt{\( (3 \times 3 \,\, \text{kernel}, \text{stride} \,\, 1, \text{padding} \,\, 1) \)} and a sigmoid activation function to produce the reconstructed keyframes, \(\hat{\bm{x}} \in \mathbb{R}^{\mathrm{I} \times \mathrm{H}  \times \mathrm{W} \times \mathrm{C}}\), initially normalized within [0, 1]. To achieve the final keyframes, denormalization scales the sigmoid output to [0, 255], ensuring the pixel values accurately reflect the visual details of the original keyframes.

\subsection{Frame Interpolation}
\textcolor{black}{Once keyframes are reconstructed, the frame interpolation module generates the complete video $\tilde{\bm{x}}$ by interpolating intermediate frames between these keyframes $\hat{\bm{x}}$. Building upon \cite{zhang_2023_CVPR}, we introduce a lightweight frame interpolation module that involves three stages: low-level feature extraction, motion and appearance estimation, and a fusion stage that outputs the color frames, as shown in Fig. \ref{Fig:semantic_decoder}. The details are presented below.}

\subsubsection{Low-level Feature Extraction}
To capture and enhance details at different scales, the low-level feature extractor uses hierarchical convolutional layers to extract multi-scale appearance features (\(\bm{l}_i^0, \bm{l}_i^1, \bm{l}_i^2\)) from each keyframe \(\hat{\bm{x}_i}\), where feature \(\bm{l}_i^k\) has dimensions \(\frac{\mathrm{H}}{2^k} \times \frac{\mathrm{W}}{2^k} \times 2^k \mathrm{C}\) in which \(\mathrm{C}\) increases as the spatial resolution decreases. These multi-scale features are achieved by using dilated convolutions with strides of \(2^{3-k}\) and dilation rates from 1 to \(2^{2-k}\). To enhance fine-grained information for subsequent motion-appearance estimation, we integrate the multi-scale features \(\bm{l}_i^k\) to complement cross-scale information by concatenating and fusing them using a linear layer to generate the cross-scale appearance feature of the \(i\)-th keyframe \(\hat{\bm{x}_i}\). Afterward, these cross-scale feature representations of keyframes are fed into the hierarchical motion-appearance feature extractor to extract both motion features and inter-frame appearance features.

 \begin{figure}[t]
    \centering
    \includegraphics[width =  0.5\textwidth]{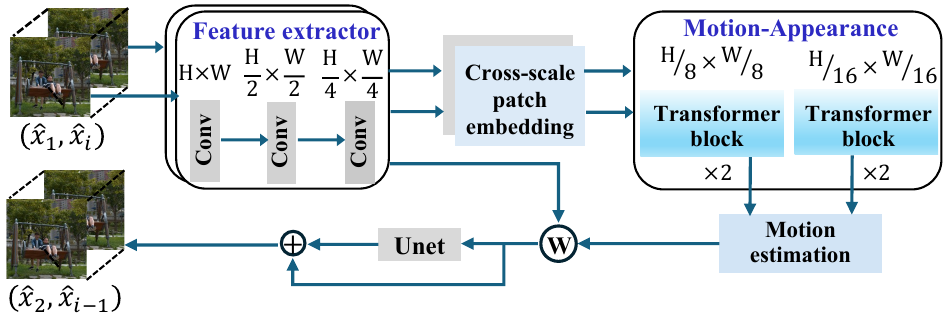}
    \caption{\textcolor{black}{Frame interpolation. }}
\label{Fig:semantic_decoder}
\end{figure}
\subsubsection{Motion and Appearance Estimation} 
Effective video frame interpolation relies on accurately capturing the motion between input frames and seamlessly integrating inter-frame appearance features, such as color and texture. To achieve this, we use inter-frame attention to naturally extract both motion and appearance details from the keyframes.

Given two keyframes, \(\hat{\bm{x}}_i\) and \(\hat{\bm{x}}_j\), we denote their appearance features as \(\bm{A}_i\) and \(\bm{A}_j \in \mathbb{R}^{\widehat{\mathrm{H}} \times \widehat{\mathrm{W}} \times \mathrm{C}}\), respectively. For a specific region \(\bm{A}_i^{m,n} \in \mathbb{R}^C\) in \(\hat{\bm{x}}_i\) and its corresponding spatial neighbors \(\bm{A}_j^{w_{m,n}} \in \mathbb{R}^{\mathrm{W} \times \mathrm{W} \times \mathrm{C}}\) in \(\hat{\bm{x}}_j\) (with \(W\) being the neighborhood window size), we generate the query \(\bm{Q}_i^{m,n}\), keys \(\bm{K}_j^{w_{m,n}}\), and values \(\bm{V}_j^{w_{m,n}}\) using 

\begin{align}
    &\bm{Q}_i^{m,n} = \bm{A}_i^{m,n} \bm{M}_Q, \nonumber\\
    &\bm{K}_j^{w_{m,n}} = \bm{A}_j^{w_{m,n}} \bm{M}_K, \nonumber\\ 
    &\bm{V}_j^{w_{m,n}} = \bm{A}_j^{w_{m,n}} \bm{M}_{V},
\end{align}
where \(\bm{M}_Q, \bm{M}_K, \bm{M}_V \in \mathbb{R}^{\mathrm{C} \times \widehat{\mathrm{C}}}\) are the linear projection matrices.

To simultaneously capture appearance information and identify motion details between the keyframes, we compute the attention map \(\bm{S}_{i,j}^{m,n}\) using the \texttt{softmax} function to measure similarity following
\begin{align}
    \bm{S}_{i,j}^{m,n} = \texttt{softmax} \left( \frac{\bm{Q}_i^{m,n} (\bm{K}_j^{w_{m,n}})^\mathrm{T}}{\sqrt{\hat{C}}} \right),
\end{align}
where \(\bm{S}_{i,j}^{m,n}\) is crucial for blending the appearance information of the two keyframes, enabling a richer understanding of how appearance transforms between frames. Specifically, we refine \(\bm{A}_i^{m,n}\) by integrating it with the weighted features from \(\hat{\bm{x}}_j\) as 
\begin{align}
    \tilde{\bm{A}}_i^{m,n} = \bm{A}_i^{m,n} + \bm{S}_{i,j}^{m,n}\bm{V}_j^{w_{m,n}},
\end{align}
where \(\tilde{\bm{A}}_i^{m,n}\) combines appearance information from both frames, which is essential for generating intermediate frames.

For accurate interpolation, we estimate the motion vector \(\bm{M}_{i,j}^{m,n} \in \mathbb{R}^{2}\) by weighting the coordinates as 
\begin{align}
    \bm{M}_{i,j}^{m,n} = \bm{S}_{i,j}^{m,n} \bm{B}^{w_{m,n}} - \bm{B}^{m,n},
\end{align}
where \(\bm{B} \in \mathbb{R}^{\widehat{\text{H}} \times \widehat{\text{W}} \times 2}\) represents a coordinate map. 

Benefiting from the timestep-invariant nature of \(\tilde{\bm{A}}_i^{m,n}\), the motion vector $\bm{M}_{i,j}^{m,n}$ can guiding subsequent motion estimation for arbitrary timestep frame predictions. Assuming local linear motion, the motion vector for an intermediate frame \(\hat{\bm{x}}_{i+\Delta}, 0 < \Delta < j-i \) is approximated as
\begin{align}
    \bm{M}_{i+\Delta}^{m,n} = \frac{\Delta}{j-i} \times \bm{M}_{i,j}^{m,n},
\end{align}
In essence, computing inter-frame attention once facilitates efficient and effective motion and appearance estimation across multiple arbitrary timestep frame predictions.

\subsubsection{Fusion}
We begin by estimating bidirectional optical flows \(\bm{F}\) and masks \(\bm{O}\) using the acquired motion and appearance features. These are then used to warp keyframes \(\hat{\bm{x}}_i\) and  \(\hat{\bm{x}}_j\) to the target frame \(i+\Delta\) and fuse them together as 
\begin{align}
    \hat{\bm{x}}_{i+\Delta} = &\bm{O} \odot \texttt{BW}\left(\hat{\bm{x}}_i, \bm{F}_{\Delta \rightarrow 0}\right) \nonumber\\
    &+ (1-\bm{O}) \odot \texttt{BW}\left(\hat{\bm{x}}_j, \bm{F}_{\Delta \rightarrow j-i}\right),
\end{align}
where \(\texttt{BW} (\cdot)\) denotes the backward warp operation.

Subsequently, we refine the appearance of the fused frame \(\bm{\hat{x}}_{i+\Delta}\) using low-level features and inter-frame appearance as
\begin{align}
    \hat{\bm{x}}_{i+\Delta} = \hat{\bm{x}}_{i+\Delta} + \text{RefineNet}\left(\hat{\bm{x}}_{i+\Delta}, \bm{l}, \bm{A}\right),
\end{align}
where RefineNet utilizes three convolution layers for motion estimation and a simplified U-Net architecture \cite{Unet} for achieving high performance.

\section{Training and Implementation}\label{training}
The proposed SD-based GSC architecture can be effectively trained via an end-to-end approach; however, this method may suffer from slow convergence. To mitigate this issue, we employ a step-by-step training strategy that systematically trains each component of the SD-based GSC architecture.

\begin{algorithm}[t]
\caption{Training Process of Semantic Denoiser}
\label{alg:semantic-denoiser}
\begin{algorithmic}[1]
\REQUIRE Training dataset $\bf{X}$, time steps $T$, hyperparameters $\left\{\beta_1, \cdots, \beta_T\right\}$, $\zeta_t^\theta$, $\zeta_t^\vartheta$, $\alpha$ and $\phi$.
\ENSURE Semantic denoisers $\zeta_t^\theta$, $\zeta_t^\vartheta$.

\STATE Train the semantic extraction and reconstruction modules using the loss function in (\ref{eq:aeloss}).

\tcp{\textbf{\textit{Training SD denoiser}}}
\STATE Freeze parameters of semantic extraction module $\mathcal{E}_{\texttt{se}}(\cdot)$.
\REPEAT
    \STATE Sample a latent representation $\bm{z}_1 \sim p(\bm{z_1})$.
    \STATE $t \sim \textrm{Uniform}(\{1,2,\ldots, T\})$, $\epsilon \sim \mathcal{N}(0, \mathbf{I})$.
    \STATE Take gradient descent step on $\nabla_{\theta} \left\|\epsilon_{\theta}(\bm{z}_1^t, t) - \epsilon \right\|_{2}^{2}$.
\UNTIL Converged

\tcp{\textbf{\textit{Denoising of SD denoiser}}}
\STATE Sample $\bm{z}_1^T \sim \mathcal{N}(0, \mathbf{I})$
\FOR{denoising step $t = T, \ldots, 1$}
    \STATE $\hat{\epsilon} \leftarrow \epsilon(\bm{z}_1^t, t)$.
    \STATE Estimate $\hat{\bm{z}}_1^t$ using (\ref{eq:z0}).
    \STATE $g \leftarrow \nabla_{\bm{z}_1^t} \log p(\bm{z}'_1|\bm{z}_1^t,\bm{h})$.
    \STATE Compute the conditional score $\bm{s} \leftarrow \zeta_t^\theta g - \frac{1}{\sqrt{1-\bar{\alpha}_{t}}} \hat{\epsilon}$. 
    \STATE Sample $\varphi \sim \mathcal{N}(0, \mathbf{I})$. 
    \STATE Compute $\bm{z}_1^{t-1} =  \frac{1}{\sqrt{\alpha_{t}} }(\bm{z}_1^{t} + \beta_t \bm{s}) + \sqrt{\beta_t} \varphi$.
\ENDFOR

\tcp{\textbf{\textit{Training PSD denoiser}}}
\STATE Train denoising estimators $\epsilon_\theta$ and $\epsilon_{\vartheta}$ for $\bm{z}_1$ and $\bm{h}$ following steps 3-7, respectively.

\tcp{\textbf{\textit{Denoising of PSD denoiser}}}
\STATE Sample $\bm{z}_1^T, \bm{h}_T \sim \mathcal{N}(0, \mathbf{I})$.
\FOR{denoising step $t = T, \ldots, 1$}
    \STATE $\hat{\epsilon}_\theta \leftarrow \epsilon_\theta(\bm{z}_1^t, t)$, $\hat{\epsilon}_\vartheta \leftarrow \epsilon_\vartheta (\bm{z}_1^t, t)$.
    \STATE Estimate $\hat{\bm{z}}_1^t$ and $\hat{\bm{h}}_t$ using (\ref{eq:z0}) and (\ref{eq:h0}).
    \STATE $g_\theta \leftarrow \nabla_{\bm{z}_t} \log p(\bm{z}'_1|\bm{z}_1^t,\bm{h}_t)$, $g_\vartheta \leftarrow \nabla_{\bm{h}_t} \log p(\bm{z}'_1|\bm{z}_1^t,\bm{h}_t)$.
    \STATE Compute the conditional score \\
        $\small \bm{s}_\theta \leftarrow\zeta_t ^\theta g_\theta - \frac{1}{\sqrt{1-\bar{\alpha}_{t}}} \hat{\epsilon}_\theta$ and $\bm{s}_\vartheta \leftarrow\zeta_t ^\vartheta g_\vartheta - \frac{1}{\sqrt{1-\bar{\alpha}_{t}}} \hat{\epsilon}_\vartheta$.
    \STATE Sample $\varphi_\theta, \varphi_\vartheta \sim \mathcal{N}(0, I)$.
    \STATE Compute $\bm{z}_1^{t-1} =  \frac{1}{\sqrt{\alpha_{t}} }(\bm{z}_1^{t} + \beta_t \bm{s}_\theta) + \sqrt{\beta_t} \varphi_\theta$ and $\bm{h}_{t-1} =  \frac{1}{\sqrt{\alpha_{t}} }(\bm{h}_{t} + \beta_t \bm{s}_\vartheta) + \sqrt{\beta_t} \varphi_\vartheta$.
    \STATE $\bm{h}_{t-1} \leftarrow \bm{h}_{t-1} - \alpha (||\bm{\bm{z}'_1} - \bm{h}_{t-1}  \bm{z}_1^{t-1}||_2 + \phi ||\bm{h}_{t-1}||)$.
\ENDFOR
\end{algorithmic}
\end{algorithm}
\subsection{Training Pipeline}
Initially, we start by jointly training the semantic extraction and semantic reconstruction modules. This phase ensures the accurate video reconstruction by integrating MSE and Kullback-Leibler (KL) divergence into the training process. The MSE facilitates accurate video reconstruction, while the KL divergence serves as a regularization term, guiding the data distribution in the latent space towards approximating a unit Gaussian distribution. The loss function of the semantic extraction and reconstruction modules is defined as the sum of MSE and KL divergence via
\begin{equation}
    \mathcal{L} = \frac{1}{N} \sum_{\bm{x} \in \bf{X}} \big|\big|\bm{x} - \tilde{\bm{x}}\big|\big|^2  + \lambda \mathcal{L}_{\text{KL}},
    \label{eq:aeloss}
\end{equation}
where $N$ is the size of dataset $\bf{X}$, and \(\lambda\) controls the significance of KL divergence relative to MSE. The KL divergence is expressed as
\begin{equation}
\mathcal{L}_{\text{KL}} = -\frac{1}{2} \sum_{i=1}^{N} (1 + \log(\sigma_i^2) - \mu_i^2 - \sigma_i^2),
\label{eq:kl}
\end{equation}
where \(\mu_i\) and \(\sigma_i\) denote the mean and standard deviation of the latent space, respectively.

Following (\ref{eq:aeloss}), we freeze the parameters of the semantic extraction module and feed the latent semantic vector \( \bm{z}_1 \) of the first frame into the semantic denoiser for training. The training and inference procedures for the \textit{SD denoiser} and \textit{PSD denoiser} under both known and unknown channels are detailed in \textbf{Algorithm 2}. For the frame interpolation module, we utilize three consecutive frames from each video sample and optimize it with the MSE loss function. This training approach ensures efficient convergence and robust performance of the SD-based GSC architecture, particularly enhancing video frame interpolation quality.

Once each component is individually trained, we finalize the process by fine-tuning the entire SD-based GSC architecture in an end-to-end manner. This fine-tuning stage optimizes the collaboration between all modules, resolving any discrepancies that may arise from independently trained components. The details of the fine-tuning process is presented in \textbf{Algorithm 3}.

\subsection{Complexity Analysis}
To evaluate the efficiency of our proposed SD-based GSC architecture, we analyze the computational complexity of each module by measuring the computational operations at each layer, as detailed in \cite{molchanov2017pruning} and \cite{LI2022109150}. This comprehensive analysis of time and space complexities provides a detailed understanding of the computational demands of each module within our architecture. Specifically, in the semantic encoder, both the time and space complexities of the semantic extraction module are \(\mathcal{O}(\mathrm{F}\mathrm{H}^2\mathrm{W}^2)\) and that of keyframe selection module are \(\mathcal{O}(\mathrm{F}^2)\); in the semantic denoiser, both the time and space complexities are \(\mathcal{O}(\mathrm{T}\mathrm{H}^2\mathrm{W}^2)\); in the semantic decoder, both the time and space complexities for the semantic reconstruction module are \(\mathcal{O}(\mathrm{I}\mathrm{H}^2\mathrm{W}^2\mathrm{C}^2)\), and that of the frame interpolation module are \(\mathcal{O}\big((\mathrm{F-I})\mathrm{H}^2\mathrm{W}^2\big)\). Given that the color channel $\mathrm{C}=3$ and  $\mathrm{T} \gg \mathrm{F}$, the overall time complexity and space complexity of inference process are the dominant term  $\mathcal{O}(\mathrm{T}\mathrm{H}^2\mathrm{W}^2)$. 
Regarding the training process, the forward and backward propagation processes effectively doubles the computational cost, but the overall complexity remains $\mathcal{O}(\mathrm{T}\mathrm{H}^2\mathrm{W}^2)$ since backward propagation does not exceed the complexity of forward propagation.

\begin{algorithm}[t]
\caption{Training Pipeline}
\label{alg:training-pipeline}
\begin{algorithmic}[1]
\REQUIRE Training dataset $\bf{X}$, time steps $T$, hyperparameters $\left\{\beta_1, \cdots, \beta_T\right\}$, $\zeta_t^\theta$, $\zeta_t^\vartheta$, $\phi$, latency requirement $T_\texttt{max}$, bandwidth $B$, transmission power $p$, channel gain $h$, noise power $\sigma^2$.
\ENSURE Reconstructed video $\Tilde{\bm{x}}$.

\tcp{\textbf{Training process}}
\STATE Jointly train the semantic extraction and reconstruction modules using the loss function in (\ref{eq:aeloss}).
\STATE Train the \textit{SD denoiser} and \textit{PSD denoiser} using Alg. 2.
\STATE Sample 3 consecutive frames from each video  to train the frame interpolation module using MSE loss.
\tcp{\textbf{Fine-tuning process}}

\STATE Freeze the parameters of the semantic extraction module and semantic denoisers.
\REPEAT
    \FOR{each sample $\bm{x} \in \bf{X}$}
        \STATE Pass through the semantic extraction module to obtain the latent representation $\bm{x}'$.
        \STATE Select keyframes $\bm{z}$ using Alg. 1.
        \STATE Use \textit{SD denoiser} to denoise the received noisy SI $\bm{z}'_1$ if $\bm{h}$ is known; otherwise, use \textit{PSD denoiser}. 
        \STATE Denoise the received noisy SI $\bm{z}'_2$ to $\bm{z}'_I$ using the channel information captured by semantic denoiser.
    \ENDFOR
    \STATE Fed the denoised data $\tilde{\bm{z}}$ to jointly finetune the semantic reconstruction module and frame interpolation module using MSE loss.
\UNTIL Converged
\end{algorithmic}
\end{algorithm}

\section{Performance Evaluation}\label{Performance}
\textcolor{black}{In this section, we evaluate our proposed SD-based GSC framework for wireless video transmission through various simulations. We consider the end-to-end transmission of video tasks from a device (i.e., NVIDIA RTX 2080TI) to a server (i.e., NVIDIA A100 80GB) over fading channels. The X4K1000FPS dataset \cite{Sim_2021_ICCV}, consisting of 4,408 video clips with 65 high-resolution (4096×2160) frames each in the training dataset and 15 video clips with 8 high-resolution frames each in the test dataset, is used as the benchmark for this evaluation.
The wireless connection between the device and server operates with a bandwidth \( B = 5 \) MHz and transmission power \( p = 1 \), with the signal-to-noise ratio (SNR) ranging from 0 to 20 dB. The implementation of SD-based GSC framework is built on Ubuntu 22.04 using PyTorch. For the semantic denoiser, we set the number of time steps \(\mathrm{T} = 1000\) and employ a linear variance scheme to determine the hyperparameters \(\{\beta_1, \beta_2, \ldots, \beta_\mathrm{T}\}\). To align with the principle of SD models for image-to-image tasks, we maintain consistency between training and inference steps to ensure high-quality image generation \cite{Rombach_2022_CVPR}. The learning rate is initialized at 0.001, and stochastic gradient descent (SGD) is used as the optimizer for the loss function.}

\textcolor{black}{
\subsection{Baselines}
To demonstrate the effectiveness of our SD-based GSC framework, we perform a comparative analysis against the state-of-the-art baselines:
\begin{itemize}
    \item \textbf{ADJSCC} \cite{Xu_2022_TCSVT}: originally designed for image transmission, ADJSCC integrates SNR adaptability into a JSCC framework to dynamically adjust to changing channel conditions by jointly optimizing encoder and decoder parameters. For video transmission, we adapted ADJSCC by processing each video frame as an independent image and applying the framework on a frame-by-frame basis for encoding and transmission.
    \item \textbf{Latent-Diff DNSC}\cite{Xu_2023_Globecom}: a JSCC-based image transmission scheme that incorporates the proposed DDPM-based semantic denoising module into JSCC framework. Similar to ADJSCC, we adapted this method for video transmission by treating each frame independently.
    \item \textbf{DeepWiVe}  \cite{Deepwive2022jsac}: divides video frames into fixed Groups of Pictures (GOPs) with four frames each. The first frame in each GOP is transmitted as a keyframe using JSCC, while the remaining frames transmit motion-compensated differences relative to adjacent keyframes, also using JSCC with distinct parameters.
    \item \textbf{DVST} \cite{Wang2023JSAC}: employs a comparable GOP structure as DeepWiVe, integrating a temporal adaptive entropy model with an Artificial Neural Network (ANN)-based nonlinear transform and conditional coding architecture to extract SI from video frames.
    \end{itemize}}
\subsection{Metrics}
We evaluate performance using the following metrics:
\begin{itemize}
    \item \textbf{MSE}: Quantifies the average squared difference between the estimated and real values.
    \item \textbf{Peak Signal-to-Noise Ratio (PSNR)}: Evaluates image reconstruction quality, with higher values indicating better performance.
    \item \textbf{Fréchet Inception Distance (FID)}: Assesses the similarity between real and reconstructed image distributions, with lower values signify better quality.
    \item \textbf{Fréchet Video Distance (FVD)}: Measures the overall quality of reconstructed videos by measuring the distance between real and reconstructed video distributions, with lower values indicating superior reconstruction.
    
\end{itemize}

\begin{figure}
    \centering
    \includegraphics[width= 0.5 \textwidth]{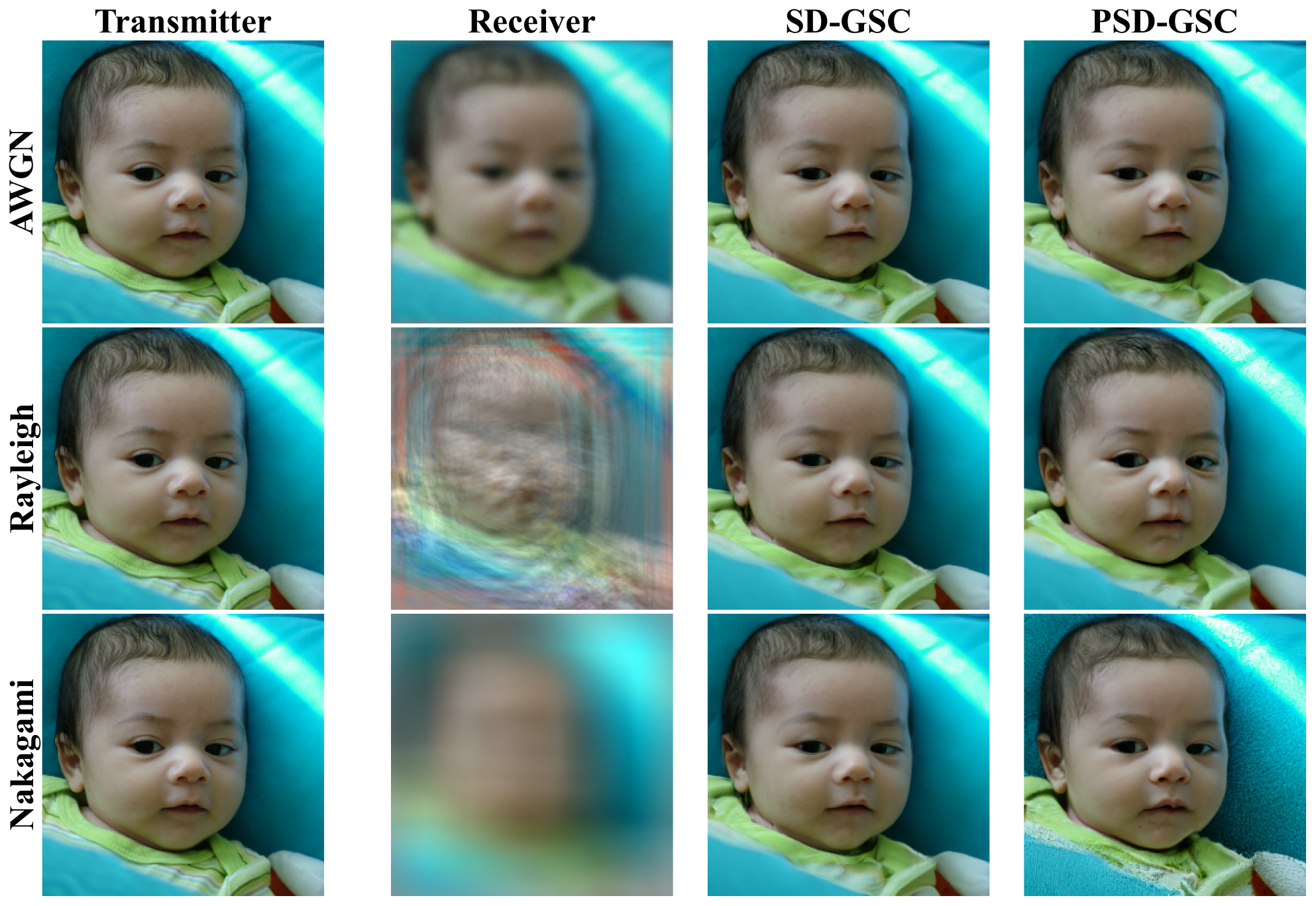}
    \caption{Example of SD denoiser-based image transmission.}
    \label{fig:subfigure}
\end{figure}
\begin{figure*}[t]
    \centering
    \subfigure[MSE]{
    \includegraphics[width=0.315 \textwidth]{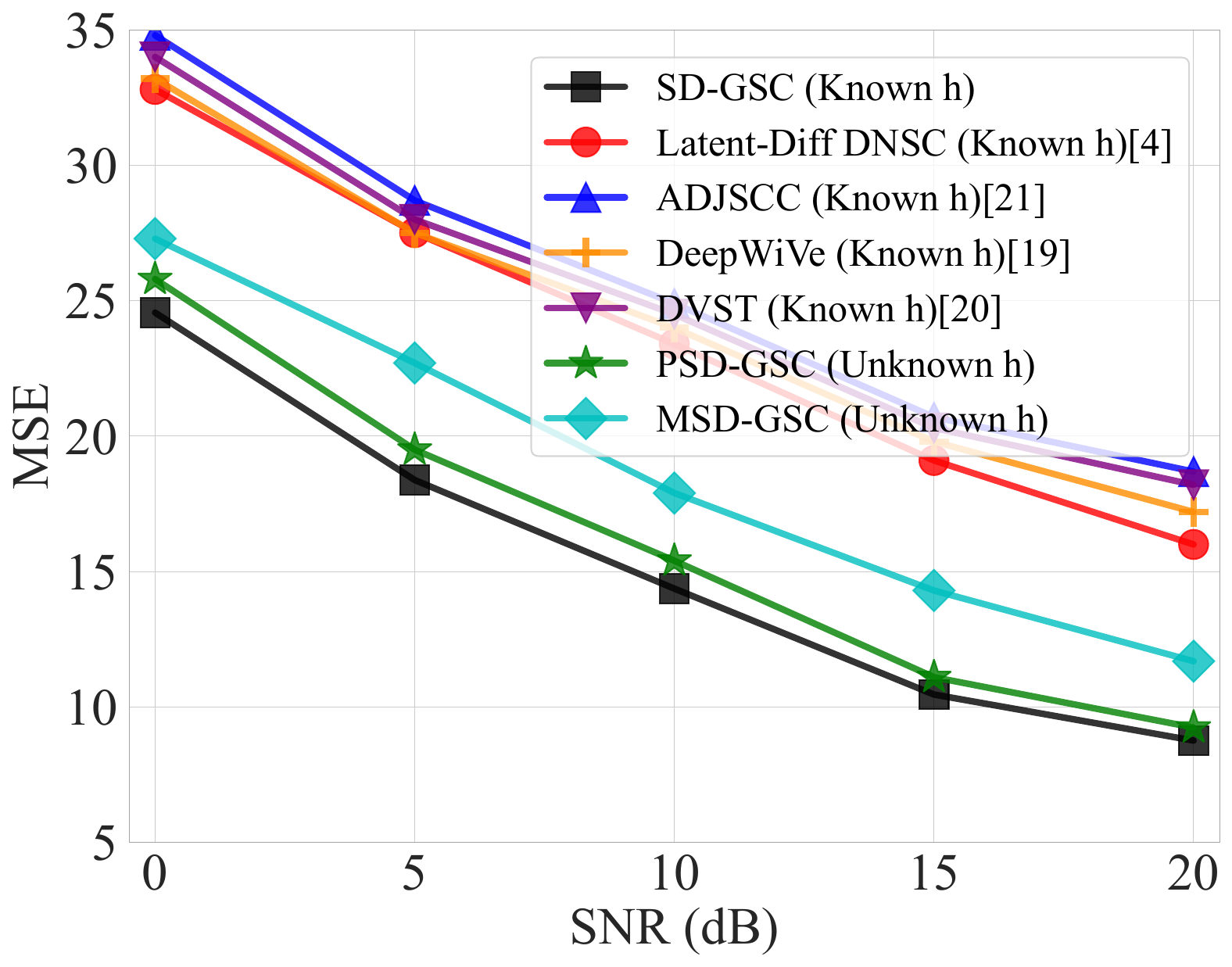}}
    \subfigure[PSNR]{
    \includegraphics[width=0.315 \textwidth]{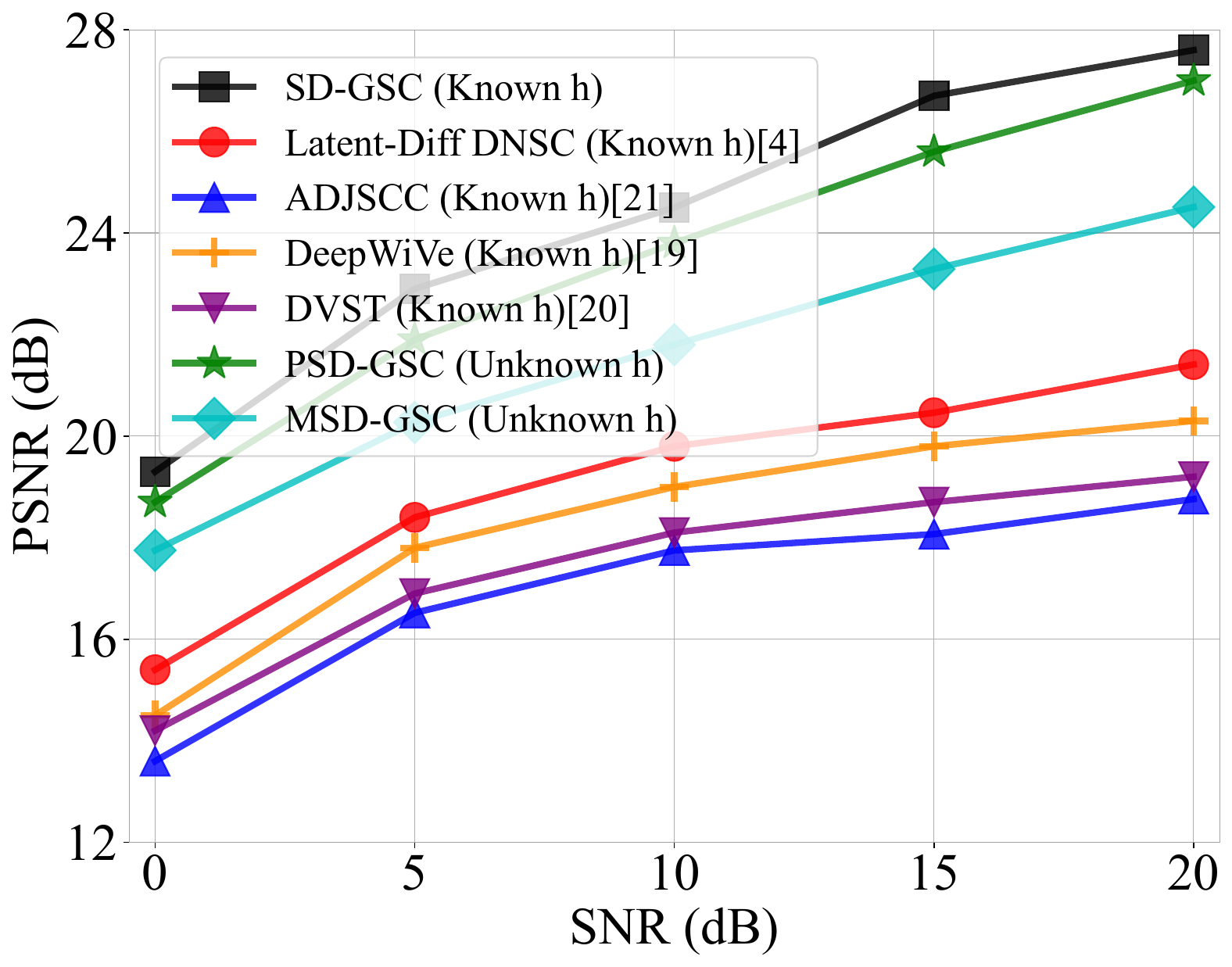}}
    \subfigure[FID]{
    \includegraphics[width=0.315 \textwidth]{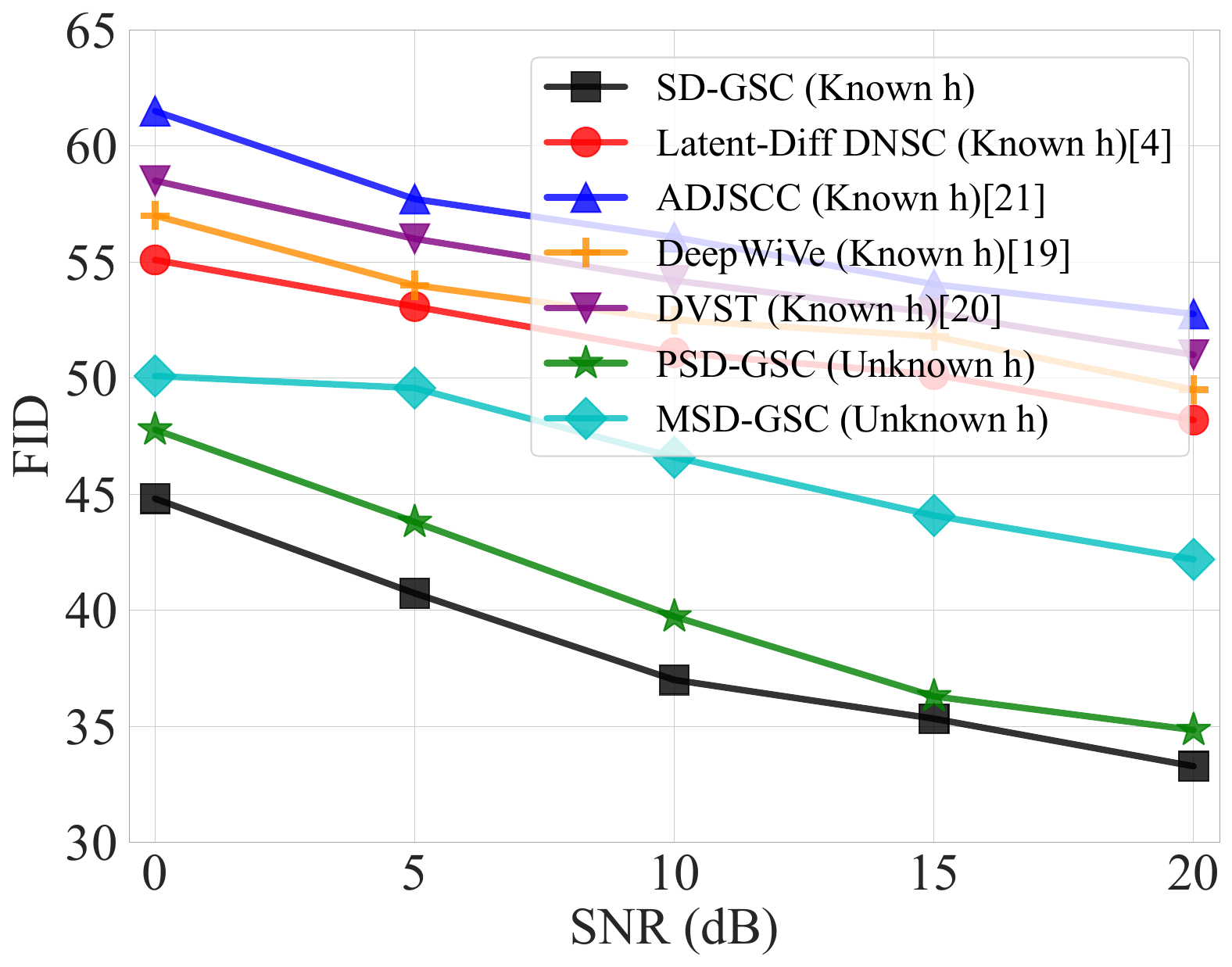}}
    \caption{\textcolor{black}{The performance of the generated first frame.}}
    \label{fig:knownH}
\end{figure*}

\subsection{Denoising capacity of SD Denoiser}
To evaluate the denoising capacity of our proposed SD denoiser and PSD denoiser, we assessed their performance for wireless image transmission under AWGN, Rayleigh fading, and Nakagami-m channels, respectively. We present one image received and recovered at the receiver under three channels for both known channel and unknown channel scenarios in Fig. \ref{fig:subfigure}. We can see that 
the quality of the denoised image remains similar across different channels when the channel gain is known. However, there is a slight deterioration observed when the channel gain is unknown, particularly noticeable under the Nakagami-m channel. This degradation, manifesting as more blurred backgrounds and subtle distortions in fine details (e.g., the baby's eyes), is due to the increased complexity of channel estimation especially in Nakagami-m channel. Despite some distortion, the overall performance remains robust, highlighting the effectiveness of our SD denoiser and PSD denoiser under varying channels.

\textcolor{black}{
Fig. \ref{fig:knownH} evaluates the performance of the first frames from each video across various SNRs under both known and unknown Rayleigh fading channels. As expected, all three metrics improve as the SNR increases, since better wireless channel conditions facilitate higher-quality image reconstruction. Interestingly, under known channels, SD-GSC and Latent-Diff DNSC outperform ADJSCC, DeepWiVe and DVST in all metrics, due to the diffusion model's ability in capturing wireless channel characteristics and removing noise effectively, resulting in more accurate reconstruction. Notably, SD-GSC outperforms both baselines in all three metrics. Specifically, SD-GSC achieves approximately 53\%, 45\%, 49\% and 52\% MSE reduction, 47\%, 29\%, 36\% and 44\% PSNR improvement,  and 37\%, 31\%, 33\% and 35\% FID reduction as compared to ADJSCC, Latent-Diff DNSC, DeepWiVe and DVST, respectively. This superior performance arises from the integration of the instantaneous channel gain, which enables better characterization of channel gain and mitigation of distortion caused by wirless channel noise, resulting in a more controlled and guided image generation than the uncontrolled generation process of DDPM used in Latent-Diff DNSC.} 

Under unknown channels, we compare our proposed PSD-GSC with MSD-GSC to validate the effectiveness of our proposed PSD-GSC for channel estimation as well as efficient image transmission. We can see both PSD-GSC and MSD-GSC experience degradation in all metrics compared to SD-GSC with known channel gain $h$. This is because channel estimation errors result in inaccurate channel gain inputs of the diffusion model, which decreases its denoising and high-quality image reconstruction capabilities. Importantly, PSD-GSC outperforms MSD-GSC in all metrics, reducing MSE by approximately 21\%, improving PSNR by 10\% and reducing FID by 18\%, respectively. This is due to the probabilistic modeling-based diffusion model used in PSD-GSC can well estimate nonlinear channels and capture the complex properties of wireless channels, for better image reconstruction. 

\begin{figure}[t]
    \centering
    \includegraphics[width=0.5\textwidth]{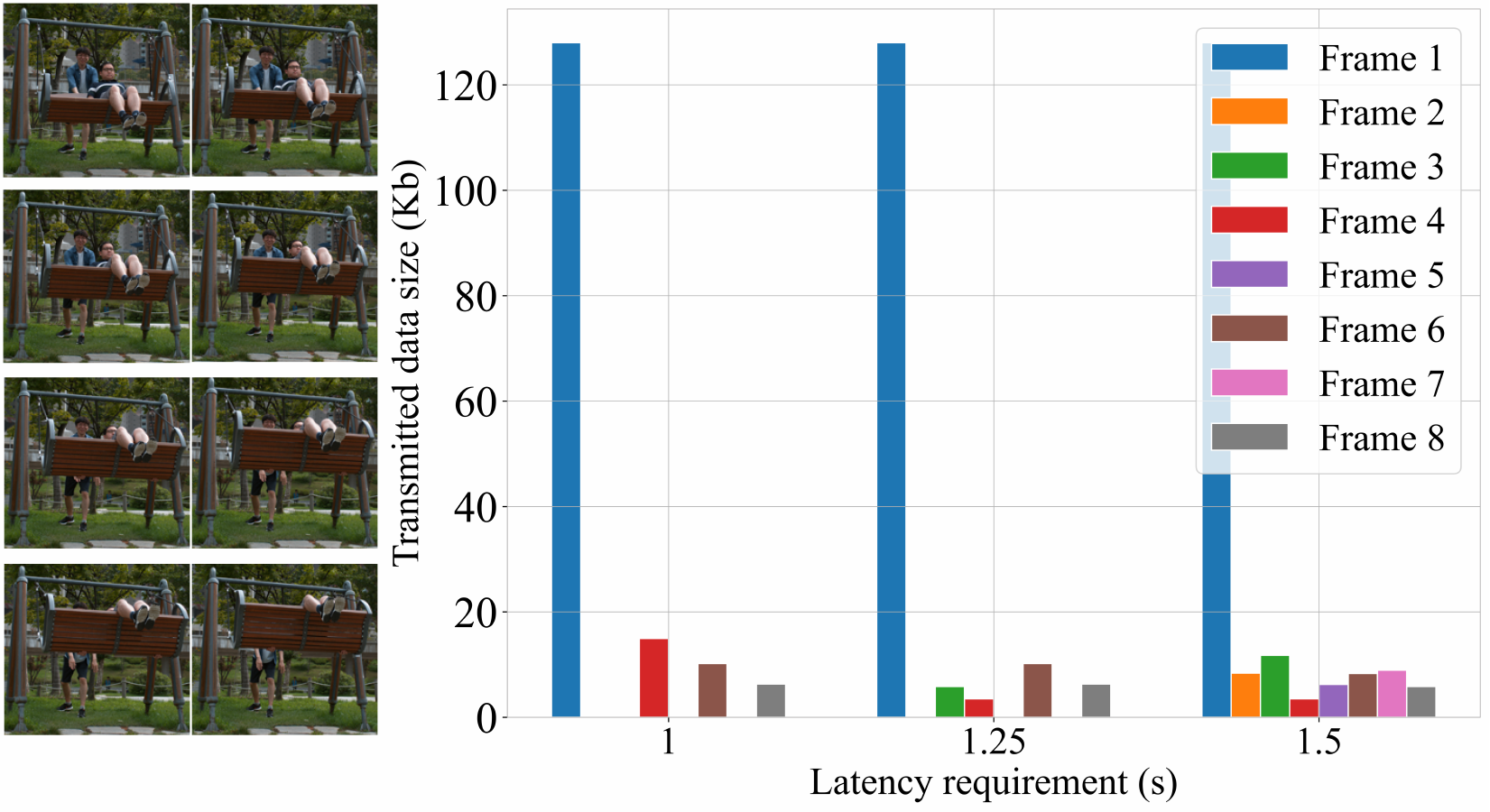}
    \caption{Keyframe selection under $\text{SNR} = 0 $ dB.}
    \label{fig:keyframe}
\end{figure}
\subsection{Video Generation}

Fig. \ref{fig:keyframe} evaluates the keyframe selection process under varying latency requirements in a known Rayleigh fading channel under the SNR of 0 dB. Under strict latency constraints, the system selects only the most critical motion changes, while more keyframes with significant motion differences are chosen as increased latency constraint. This occurs because higher latency budget enables the transmission of additional keyframes, that can capture finer motion details with improvement in video reconstruction quality. This approach effectively balances the trade-off between transmission delay and video quality, enhancing the overall performance of the semantic video communication system by optimizing bandwidth usage, ensuring transmission of goal-oriented semantic information. 

\begin{figure*}[t]
    \centering
    \subfigure[MSE]{
    \includegraphics[width=0.315 \textwidth]{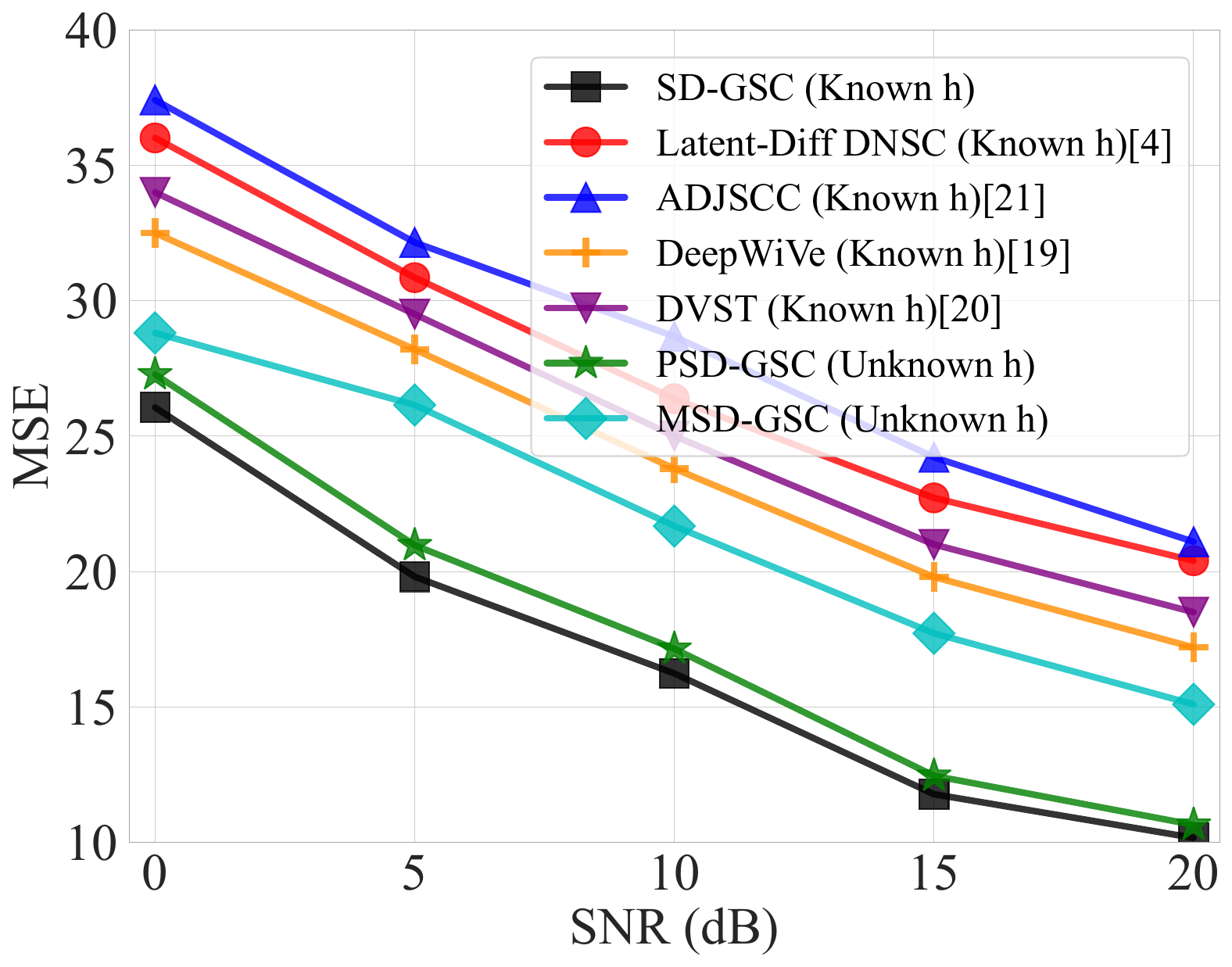}}
    \subfigure[PSNR]{
    \includegraphics[width=0.315 \textwidth]{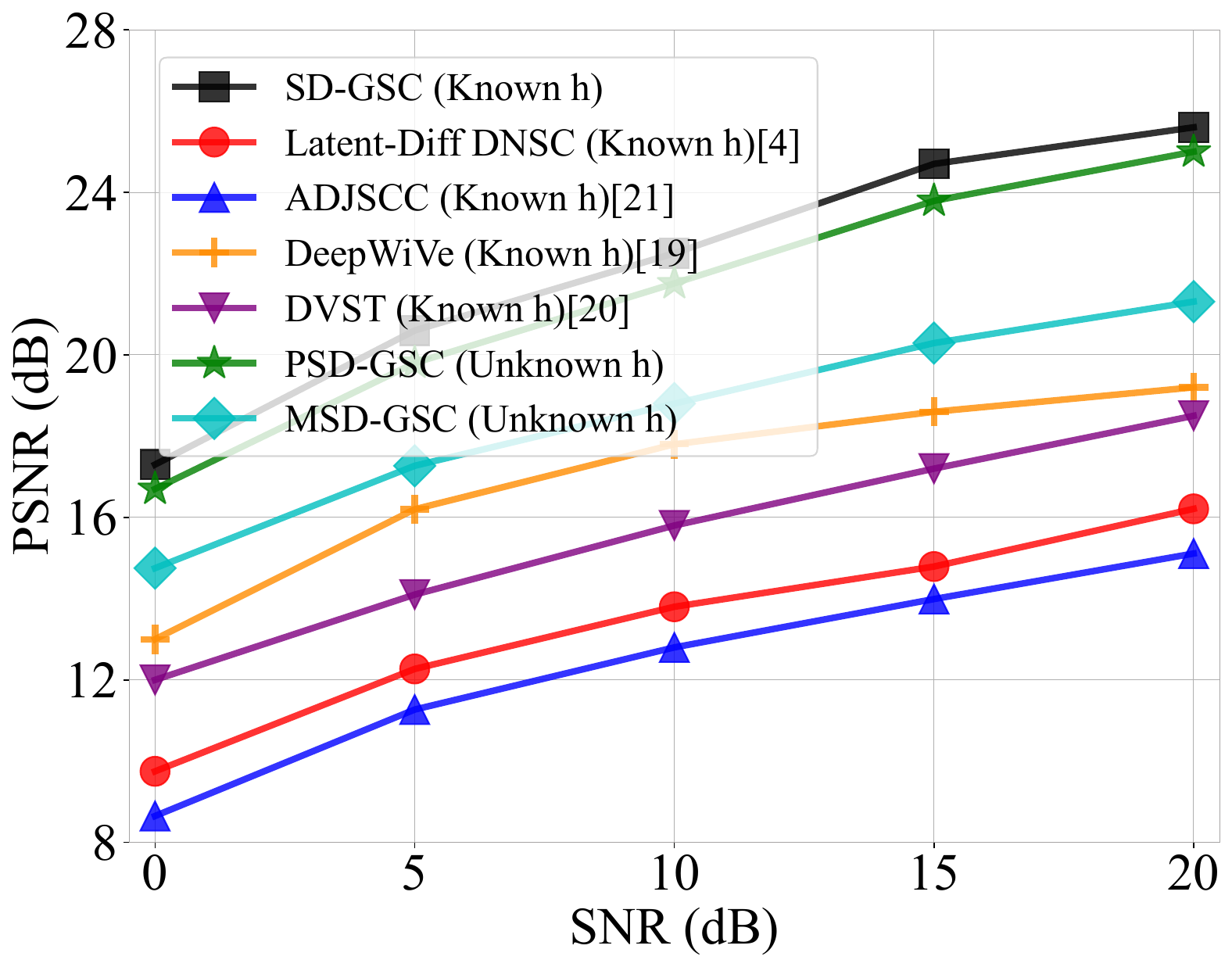}}
    \subfigure[FVD]{
    \includegraphics[width=0.315 \textwidth]{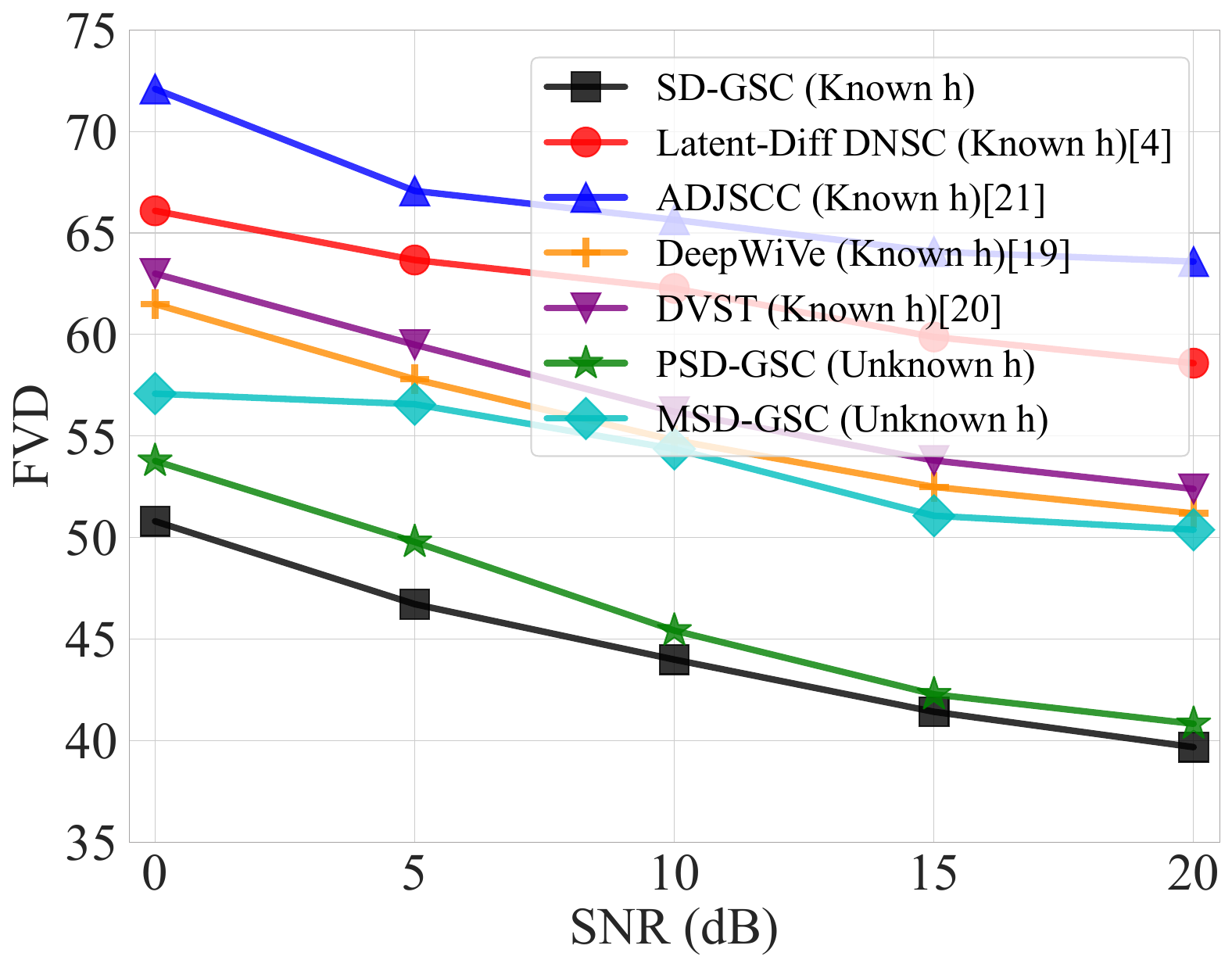}}
    \caption{The performance of the generated video frames under latency requirement of 1s.}
    \label{fig:video}
\end{figure*}

\textcolor{black}{Fig. \ref{fig:video} evaluates the performance of our proposed frameworks for wireless video transmission over a Rayleigh fading channel under a 1s latency constraint. As shown, DeepWiVe and DVST achieve better performance than Latent-Diff DNSC, despite Latent-Diff DNSC outperforming them in the first frame. This is because the frame-by-frame processing of Latent-Diff DNSC faces frame loss due to strict latency constraints, and the uncontrollable frame generation leads to degraded video generation quality. In comparison, our SD-GSC achieves significantly better performance than ADJSCC, Latent-Diff DNSC, DeepWiVe, and DVST, achieving MSE reductions of approximately 52\%, 50\%, 41\%, and 45\%, respectively; PSNR improvements of 69\%, 58\%, 33\%, and 38\%; and FVD reductions of 38\%, 32\%, 22\%, and 24\%. Additionally, PSD-GSC reduces MSE by approximately 29\%, improves PSNR by 17\%, and reduces FVD by 19\% compared to MSD-GSC. This superior performance is owing to two primary reasons. First, our framework can dynamically select keyframes and identify the differences between consecutive keyframes to reduce the transmitted data volume, subsequently reconstructing the missing frames using the frame interpolation module to improve video transmission quality. Second, the denoising capability of our proposed semantic denoiser improves the characterization of the wireless channel, thereby enhancing video generation quality.}

We can also see that both SD-GSC and PSD-GSC in Fig. \ref{fig:video} show some performance degradation in video reconstruction compared to the generated initial frame in Fig. \ref{fig:knownH}. This degradation is primarily due to reconstruction loss in the frame interpolation process, where the interpolated frames may not perfectly match the original frames. \textcolor{black}{Additional factors cause this degradation include cumulative errors in the frame interpolation process, especially for frames further away from keyframes, challenges in accurately capturing complex motion patterns between keyframes in scenes with rapid or unpredictable movements, and potential loss of fine details during the keyframe selection and interpolation processes that may not be fully recovered in the reconstructed video.} Nevertheless, our proposed framework still outperforms existing approaches, demonstrating their effectiveness in balancing video quality and latency constraints in wireless transmission scenarios.

\section{Conclusion}\label{Conclusion}
In this paper, we have developed a stable diffusion (SD)-based goal-oriented semantic communication (GSC) framework for efficient video transmission under various fading channels. The main goal is to achieve high-quality video reconstruction at the receiver under latency constraints. To this end, we have first designed a semantic encoder to effectively extract the relevant semantic information (SI) by identifying the keyframes and their semantic representation thereby reducing the transmission data size. To accurately reconstruct the video from the received semantic information, we then 
developed a semantic decoder to reconstruct the keyframes from the received SI and further generated the full video from the reconstructed keyframes using frame interpolation. To combat the negative impact of wireless channel noise on the transmitted SI, we have proposed an SD-based semantic denoiser for known channel scenario to better characterize and mitigate the channel noise, resulting in  significant improvements in both MSE, PSNR and FVD metrics compared to the state-of-the-art baselines. 
For unknown channel scenario, we have further designed a parallel SD-GSC (PSD-GSC) to jointly learn the channel gain and denoise the received SI, showcasing the capability of our proposed PSD-GSC in channel estimation and efficient video transmission. This work opens new research avenues for integrating advanced SD models in GSC to achieve high-quality transmission of video SI in dynamic and noisy wireless channel environments. Future work will focus on optimizing the application of SD models in GSC to further improve the robustness of multimodal data transmission.

\bibliographystyle{IEEEtran}
\bibliography{ddpm}

\vfill

\end{document}